\documentclass[journal=mamobx,manuscript=article]{achemso} 
\setkeys{acs}{articletitle=true,etalmode=truncate,maxauthors=0}

\usepackage{titlesec}
\usepackage{graphicx}
\usepackage{dcolumn}
\usepackage{bm}
\usepackage{float}
\usepackage{amsmath}
\usepackage{amssymb}
\usepackage{color}
\usepackage[font=footnotesize]{caption}
\usepackage{gensymb}
\usepackage{verbatim}
\usepackage[switch]{lineno}
\usepackage{hyperref}
\usepackage[title]{appendix}
\usepackage{booktabs}

\hypersetup{colorlinks=true, citecolor=blue, urlcolor=blue, linkcolor=blue}
\titleformat{\section}[block]{\normalsize\bfseries}{\Roman{section}}{0.4em}{\MakeUppercase}
\titleformat{\subsection}[block]{\normalsize\bfseries}{\Roman{section}.\Alph{subsection}}{0.4em}{}

\author{Jian Li}
\affiliation[UoH]{Department of Physics and Electronic Engineering, Heze University, Heze 274015, China}
\author{Bokai Zhang}
\email{zbk329@swu.edu.cn}
\affiliation[UoS]{School of Physical Science and Technology, Southwest University, Chongqing 400715, China}
\affiliation[ZSTU]{Department of Physics, Zhejiang Sci-Tech University, Hangzhou 310018, China}
\author{Yushan Li}
\email{lysh507@163.com}
\affiliation[UoH]{Department of Physics and Electronic Engineering, Heze University, Heze 274015, China}

\title[\texttt{achemso} demonstration]{Glass formation in mechanically interlocked ring polymers: the role of induced chain stiffness}

\begin{document}
\begin{abstract}
Polymer-related materials exhibit rich glassy behaviors at different length scales due to their various molecular structures and topological constraints. Recent studies have identified transient interpenetration of the long-chain rings contributing to dynamic arrest on the center-of-mass level.  Interpenetration of rings is proposed as an approach to facilitate glass formation in polymer melts.  In this work, inspired by recent advances in the synthesis of mechanically interlocked polymers, we investigate glass transition on the nanometer-scale segments influenced by permanent interpenetration of rings using molecular dynamics simulations.  We find that decreasing chain length in the mechanically interlocked system is equivalent to inducing an effective chain stiffness on the sub-rings.  The induced stiffness provides a unified explanation for these unique structural features and transient dynamic arrest in the system of interlocked rings with rather short chains.  Further, a crossover is observed in the scaling relation between localization and glassy depth upon cooling.  Our work reveals a dynamic transition from weak to strong caging at the crossover temperature.  According to the localization model, we demonstrate that the chain stiffness increases the critical temperature and oscillation distance, therefore leads to more fragile dynamics and deeper glassy state.  These findings are consistent with the predictions of molecular simulations and theories for polymers with real local stiffness.  Our work deepens the understanding of the role of induced stiffness on glass transition, and opens up a new direction to design rich glass materials by manipulating stiffness through mechanical bonds. 
\end{abstract}
\noindent\rule{\linewidth}{1pt}

\section{Introduction}
Understanding glass transition in polymer-related materials is a yet unsolved issue in polymer physics due to their diverse architectures and topological constraints, which are ubiquitous in living organism and industrial synthesis \cite{Rosa2019,Rosa2008,Tezuka2013}.  Molecular internal structures and their topological constraints have been found to give rise to many unique structural, dynamic and rheological properties in linear chains \cite{Doi1988,Rubinstein1985,Larson2005,Ralf2004,Tzoumanekas2006}, rings \cite{Kapnistos2008,Halverson2011Static,Halverson2011Dynamics,Leopoldo2020} and nanocomposites \cite{LiYing2012,Chen2019,Cao2019,Zhang2021}. Further, much effort has been devoted to the study of glass-forming polymers driven by different topological interactions, including transient interpenetration in rings \cite{Moreno2014}, intrachain knot complexities \cite{Floudas2018,JCP_Douglas_2019}, and chemically linking in giant molecular clusters \cite{Stephen2019, Liu2022}.

For dense solutions of unknotted and nonconcatenated rings, the concept of a novel topological glass has recently been proposed and developed \cite{Lo_2013,Moreno2014,Michieletto2016,Michieletto2017,Orlandini_2021,Smrek2020,Smrek2020b,Smrek2022}.  Interpenetrating rings form amorphous cluster phase and generate extensive networks, which result in the center-of-mass slow relaxation featuring soft caging and decouple of self-collective correlations \cite{Moreno2014}.  Rings in the transient cluster only can be relaxed via sequential unthreading. Their structural relaxation time exhibits an exponential increase with chain length, leading to a conjecture that topological glass would emerge in the limit of long rings \cite{Lo_2013}.  Recently, some approaches have been widely used to induce a dynamically arrested state by enhancing interpenetration of rings, e.g., randomly pinning perturbation \cite{Michieletto2016,Michieletto2017}, and adding segmental activity \cite{Smrek2020,Smrek2020b,Smrek2022}.  Randomly pinning a fraction of rings in solution, unpinned fraction exhibits apparent glassy behaviors, characterized by a decrease of the center-of-mass diffusion by about two orders of magnitude \cite{Michieletto2016}.  A universal scaling relation predicting glass transition at high density and long chain can be obtained by continuously decreasing the perturbation to zero \cite{Michieletto2017}.  Another approach is to introduce activity on some segments of rings, resulting in a significant increase in the number of interpenetrating rings.  This triggers "active glass transition", that is, a dramatic slowing of chain motion \cite{Smrek2020}.  Note that conventional glassy behaviors in well-studied polymer glasses usually occurs on the segmental scale \cite{Roth2016}.  In contrast, within the existing topologically driven glasses mentioned above, dynamically arrested state refers to the motion inhibition of the entire chains and requires quite long rings, but intrachain dynamics remain ergodic and free to relax.

Compared with these transient interpenetration of rings in the above topological glasses, recent advanced techniques for the synthesis of mechanically interlocked polymers  offer us a possibility to investigate glassy dynamics influenced by permanent interpenetration \cite{Hart2021Review,Emilio2019Review,Niu2009,Gibson2014,Weidmann1999}. Mechanically interlocked rings (MIRs) possess very different topological constraints.  The structural parameters in the MIR molecules (such as sub-ring length, knots and assembly way) strongly affect the translational and rotational motion of sub-rings.  For example, sub-rings in MIRs can move on the scale much smaller than chain size.  Some sliding motion can occur for rotaxane-based MIPs. \cite{Chen2022}  Microstructures and constrained motions could be effectively controlled by varying the cross-linking of sub-rings . \cite{Yulin2022} On larger length scale, these sub-rings cannot be separated without breaking covalent bonds.  The fundamental mechanical interlocking between two rings has been widely used to construct various MIRs materials with complex topology, such as polyrotaxanes and polycatenanes \cite{Hart2021Review}.  Previous simulations have been carried out to study the thermodynamic, static and dynamic properties induced by the mechanical bonds in polycatenanes, in which multiple mechanically interlocking rings constitute a linear molecule \cite{Rauscher2018AML,Rauscher2020MA,Rauscher2020JCP}.  In particular, the slowing down of relaxation and the broadening of its distribution have been found in an isolated polycatenane \cite{Rauscher2018AML}.  However, the above studies have primarily focused on normal liquids regime.  The study regarding structural features and dynamic properties specific to the system with mechanical bonds in supercooled liquids regime is currently lacking. Further, encouraged by the development of synthetic strategies towards mechanically interlocked oligomers \cite{Lewis2020}, we mainly focus on dense melts of the two interlocked rings with rather short chain, which are the fundamental unit of complex mechanically interlocked molecules \cite{Leigh2014, Kelly2004,Danon2017, Sauvage1989, Marcos2016}.  

In this article, we investigate glassy formation driven by the permanent interlocked mechanical bonds, which is characterized by local monomer caging and slow segmental dynamics. 
We perform extensive molecular dynamics simulations for a model polymer,  where a molecule consists of two interlocked sub-rings.  We characterize the change of sub-rings shape and overall stiffness under the influence of mechanical interlocking by structural features, and further propose their relationship with segmental dynamics. Based on the simulation results and the localization model that provides a quantitative relation between long-time structural relaxation and short-time vibrational motion \cite{Douglas2015PNAS}, this paper aims to answer the below questions : (1) what are features of sub-ring conformations unique to the system with mechanical bonds, especially for short chains?, (2) how do these chain length and interlocked mechanical bonds on chain length scale affect the monomer packing and motion?, (3) how do the mechanical bonds affect segmental dynamics, including structural relaxation, diffusion and localization? and is there a unified explanation for the effect on these physical quantities?, (4) Are there some simple theoretical perspectives to understand the relation between the long-time dynamics and other physical quantities on microscopic scale, or clarify the combined effect of chain length and the novel mechainical bond?  

\section{Model and Method}
We have performed molecular dynamics simulations (MD) of coarse-grained polymers where a molecule consists of two interlocked sub-rings (IRs) as shown in the inset of Figure \ref{fig:1}. For comparison, we also consider a system of nonconcatenated rings (NRs).  We use the well-known Kremer-Grest bead-spring model polymer \cite{Kremer1990JCP,Binder1998}.  The excluded volume interactions between all the monomers are given by the shifted Lennard–Jones potential,
\begin{eqnarray}\label{1}
U_{LJ} (r)=\left\{
\begin{array}{lll}
 4\epsilon\Big[ \Big(\frac{\sigma}{r}\Big)^{12}-\Big(\frac{\sigma}{r}\Big)^6\Big]+U_c,\ r\le r_c \\
 0, \qquad \qquad  \qquad\qquad \ \ \ \ otherwise
\end{array}
\right.
\end{eqnarray}
where $r$ is the separation distance between two monomers.  The cutoffs are respectively taken as $r_c=2r_{min}$ for nonbonded monomers, $r_c=r_{min}$ for bonded monomers, where $r_{min}=2^{1/6}\sigma$ is the location of purely repulsive LJ potential.  $U_c$ is a constant energy shift, ensuring that the potential is continuous at the cutoff.  Besides, bonded monomers connect via a finitely extensible nonlinear elastic (FENE) potential
\begin{equation}\label{2}
   U_{FENE} (r)=-\frac{K}{2}R_0^2ln\Big[ 1-\Big(\frac{r}{R_0}\Big)^2\Big] 
\end{equation}
with the maximum separation is set as $R_0=1.5\sigma$ and spring constant $K=30\epsilon/\sigma^2$.  The combined effect of the FENE potential and $U_{LJ}$ yields an effective bond potential with a sharp minimum at $r_b\approx0.9606\sigma$.  Both the bonded and excluded volume potential prevent the chains from crossing each other even at high concentration.  Therefore, the model polymers can be applied to prepare permanent mechanical bonds in a melt of interlocked rings.  In the work, our simulation results are presented in reduced LJ units.  The units of temperature, density and relaxation time are respectively $T=\epsilon/k_B$, $\rho=\sigma^{-3}$, and $\tau_{LJ}=\sqrt{m\sigma^2/\epsilon}$.  The units of energy and length are taken as $\epsilon=1$ and $\sigma=1$.   The Boltzmann constant and the mass of monomer are also set to unit, namely $k_B=1$ and $m=1$.    

We consider a system containing $M_c=1000$ molecules and thus $2\times M_c$ sub-rings.  We simulate two different systems as a comparison, nonconcatenated rings (NRs) and two interlocked rings (IRs).  Chain lengths of sub-rings are set as $M=7$, $8$,$10$, and $20$, respectively.  All the simulations are conducted using the LAMMPS software.  The simulation box is cubic.  Periodic boundary conditions are applied in all directions.  The MD integration time step is $\delta t=0.001$.  The initial configurations are prepared by randomly placing the polymers at the sites of a simple large cubic lattice.  The initial simulation cell is much larger than the typical size of polymers studied and molecules do not overlap each other.   We first perform MD simulation in the NPT ensemble with Nose-Hoover thermostat and until the volume shrinks to a fixed volume with monomer number density $ \rho= 1.05$.   The systems are then equilibrated for $10^8$ time steps under NVT condition.  Production simulation is performed under NVE ensemble.  All the quantities are averaged over 10-20 independent initial configurations.

\section{Results and discussion}
\subsection{Sub-rings conformation}
We first explore the effect of mechanical bonds on sub-ring conformations by calculating the intra-ring structure factor in the spatially homogeneous and isotropic melt,
\begin{equation}\label{3}
    \omega(q)=\frac{1}{M}\Big\langle \sum_{i,j=1}^M\frac{sin (qr_{ij})}{qr_{ij}}\Big\rangle
\end{equation}
Here $q$ is the wave vector in reciprocal space, $r_{ij}=|\bm{r}_i-\bm{r}_j|$ denotes the distance between monomer $i$ and monomer $j$ belonging to the same sub-ring with chain length $M$, and the angular brackets represent the average over all configurations of sub-rings in the melt.  

\begin{figure}[!t]
\centering
	\includegraphics[width=\linewidth]{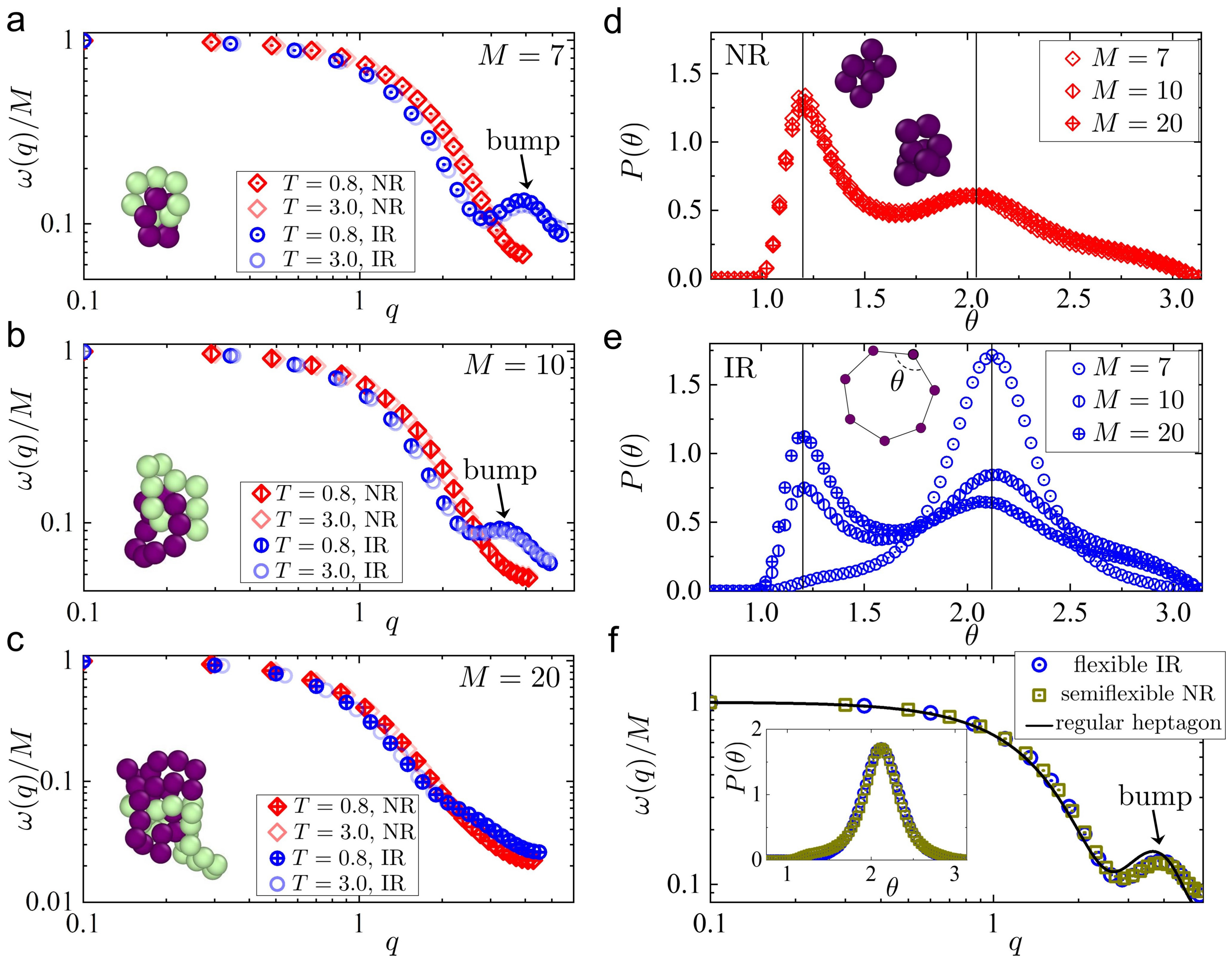}
	\caption{(a)-(c)Normalized intra-ring structure factors as a function of wave vector for nonconcatenated rings (NR, red diamonds), interlocked rings (IR, blue circles) at low temperature $T=0.8$ and high temperature $T=3.0$ (semitransparent symbols).  The data for three representative sub-ring lengths, $M=7$ (a), $M=10$ (b), $M=20$ (c), are plotted, respectively.  Inset: representative snapshots of  two interlocked rings in simulation with corresponding sub-ring lengths. (d) Bond angle distributions of NR systems with three different sub-ring lengths.  Inset: representative collapsed conformations for NRs with $M=7$ and $M=10$, respectively.  (e) Bond angle distributions for IR systems.  (f) Comprision of normalized intra-ring structure factors of flexible IR and semiflexible NR with real intramolecular rigidity at $M=7$.  The solid line represents eq \ref{4} with $M_v=7$ (regular heptagon).  Inset: bond angle distributions for the two systems of flexible IR and semiflexible NR, respectively.}
	\label{fig:1}
\end{figure}

Normalized intra-ring structure factors $\omega/M$ for the systems of NRs and IRs are shown in Figure \ref{fig:1}a-c, respectively.  For the system of IRs, a pronounced bump in the structure factor is observed on the intermediate length scale, $\sim qR_g\approx 4$.  With increasing chain length, this bump decreases and ultimately disappears at $M=20$ (Figure \ref{fig:1}c).  The intermediate-length bump is absent for NRs regardless of the chain length.  Note that the same bump has been observed in previous simulation for semiflexible ring melts \cite{Li2020}.

To understand the unique structural features belonging to short interlocked rings, we turn to calculate the probability distribution function (PDF) of bond angles to characterize the ring shape influenced by mechanical interlocking.  As shown in Figure \ref{fig:1}d, the bond angle distributions for the system of NRs with different chain lengths are almost identical.  There are two peaks at $\theta\approx1.2$ radians and $\theta\approx2.1$ radians, respectively.  The first peak is more pronounced, indicating the collapsed conformation of sub-rings (inset of Figure \ref{fig:1}d). Whereas for the system of IRs, the bond angle distributions show very different chain length dependence.  With decreasing chain length, Figure \ref{fig:1}e shows that the first peak has a significant decrease and the second peak strongly develops.  When $M=7$, a rather short ring, the first peak almost disappears.  This indicates a morphological transition from compact to expanded and tight rings.  These behaviors are very similar to that of system of semiflexible nonconcatenated rings with real local chain stiffness.

To see this, we simulate a melt of semiflexible nonconcatenated rings with $M=7$.  The intramolucular rigidity is explicitly implemented by bending energy $V_B=(\epsilon K_B/2)(cos\theta_i-cos\theta_0)^2$ with $K_B=11.2$ and $\theta_0=2.43$ radians.  These parameters ensure that the bond angle distributions of semiflexible NR and flexible IR are identical and their second peak locates the same position (the inset of Figure \ref{fig:1}f) at the same polymer concentration.  The system with real chain stiffness shares a similar intermediate-length bump to that of IRs as shown in Figure \ref{fig:1}f.  Therefore, we speculate that chain stiffness expands the rings, thereby results in the bump.  To verify this further, we calculate the intra-ring structure factor for a regular polygon by the following exact and explicit expression \cite{Li2020},
\begin{equation}\label{4}
\omega_{polygon}(q)=\frac{1}{M_v}\sum_{i,j}^{M_v}\frac{sin(qL_{ij})}{qL_{ij}}
\end{equation}
Here, $M_v$ is the number of vertices in a polygon, and $L_{ij}$ is the scalar distance between vertex $i$ and $j$, which is known for specific regular polygon.  The regular polygon can be regarded as a shape of ring when chain stiffness is infinity.    Figure \ref{fig:1}f shows the bump at intermediate length for regular heptagon ($M_v=7$) at the same bond length while it is located at slightly smaller wavevector.  Note that the bond angle of regular heptagon is 2.24 radians, which does not strictly equal to the second peak position of bond angle distribution of IR with $M=7$.  This reflects the fact that they are compressed at high polymer concentration and thus not a perfect heptagon.  Our results above demonstrate the intermediate-length bumps in $\omega(q)$ are inherently related to ring expansion.  In the system of IRs, mechanical interlocking expands the two interlocked sub-rings and reduces their accessible conformations.  This is equivalent to inducing an effective and implicit local stiffness on a chain.  

Our calculation also provides a potential explanation for the same intermediate-length bump found in dense solution of poly[7]catenanes, where 7 sub-rings are interlocked in a linear fashion and each sub-ring has 15 monomers \cite{Rauscher2020MA}.  We have also evidenced that mechanical interlocking significantly alters the bond angle distribution and chain stiffness at $M=15$, and thus leads to the bump. Besides, the bump has a slight decrease at $T=3.0$, resulting from the fact that a ring at high temperature becomes more flexible.  
\subsection{Local structures}
\begin{figure}[!t]
\centering
	\includegraphics[width=0.5\linewidth]{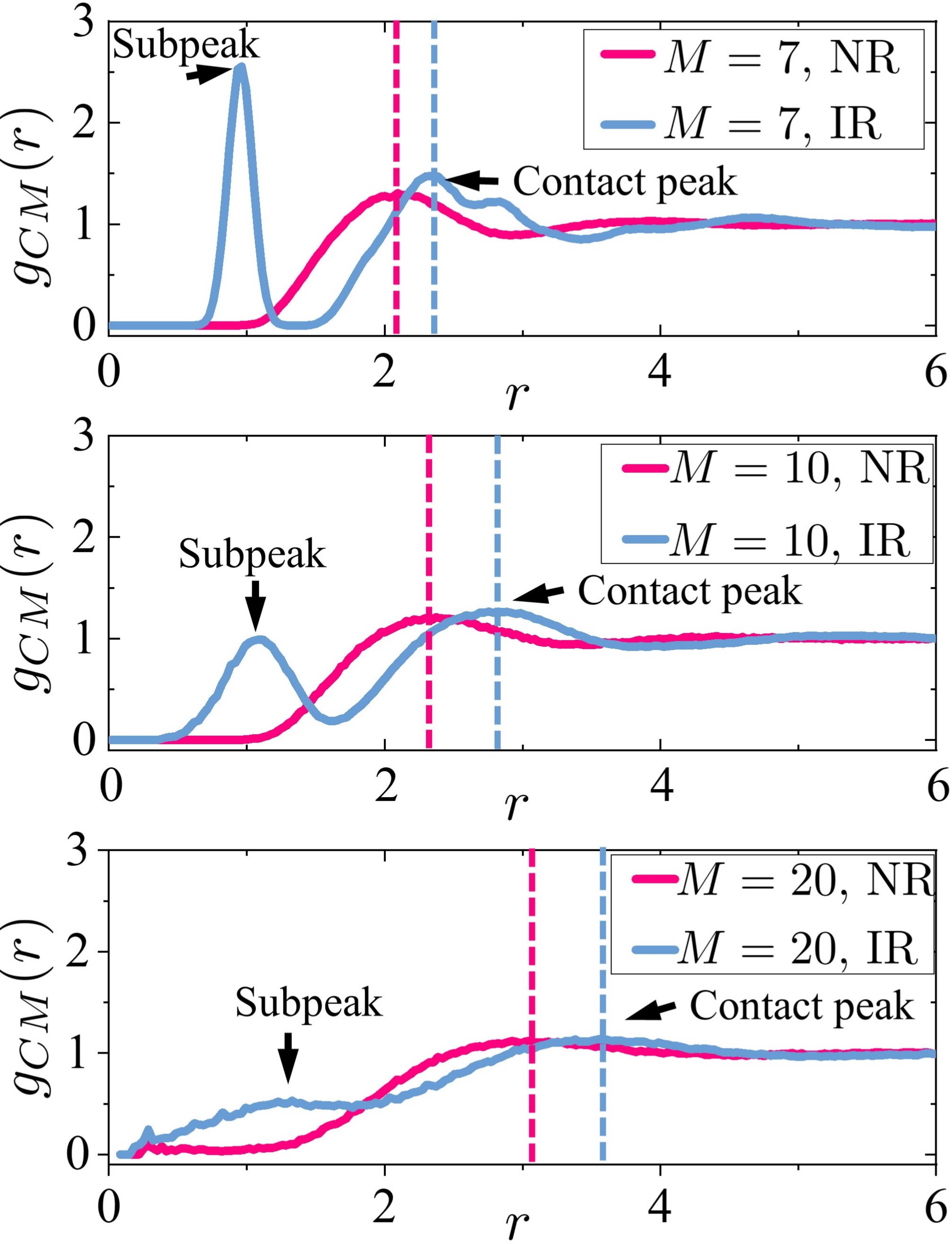}
	\caption{The center-of-mass radial distribution functions of nonconcatenated rings (NR, red line) and interlocked sub-rings (IR, blue line) on chain length $M=7$ (top), $M=10$ (middle), and $M=20$ (bottom).  The vertical dashed lines represent the position of contact peaks of the CM-RDF.  All the above data are obtained at $T=0.8$.}
\label{fig:2}
\end{figure}

To further quantify the expansion of ring induced by mechanical bonds, we consider static inter-ring correlations by calculating the center-of-mass radial distribution functions of sub-rings (CM-RDF).  As shown in Figure \ref{fig:2}, a peak is observed for both NR and IR systems at a separation of $r>2$.  This peak is called as contact peak, representing the probability of finding the nearest neighbor rings around a sub-ring that does not belong to the same molecule.  The position of the contact peak, $d_{eff}$, corresponds to a contact distance between two sub-rings.  We therefore call it effective contact diameter.  As shown in Figure \ref{fig:2} and Table \ref{tab:1}, the contact diameters of IRs are larger than that of NRs for all the chain lengths studied, revealing the expansion effect of mechanical bonds on sub-rings.  These results are consistent with the fact that IRs have larger diameters of gyration (Table \ref{tab:1}).  Besides, the decrease and broadening of the contact peak with chain length reflects the fact that the long and flexible rings behaves like a strongly overlapping ultrasoft particles \cite{Moreno2013}.  The expansion effect can be quantified as the difference between the diameters of gyration of IRs and NRs, $d_g^{IR}-d_g^{NR}$.  The quantity is found to be indepenent of chain length, $d_g^{IR}-d_g^{NR}\approx 0.3$ as shown in Table \ref{tab:1}, suggesting that the mechanical interlocking leads to a hardcore with constant diameter in the middle of sub-rings.  The characteristic ratio of polymer chains is defined as $C=(R_g^2)^{IR}/(R_g^2)^{NR}$, which is a reliable estimate of dynamic correlation or Kuhn length \cite{Ding2004}. When chain length decreases from 20 to 7, the characteristic ratio increases from 1.20 to 1.33, indicating enhanced chain stiffness. 

\begin{table}[!t]
\caption{\label{tab:1}%
Summary of the effective contact diameters, $d_{eff}$ , and the diameters of gyration, $d_{g}$, for the systems of nonconcatenated rings (NR) and interlocked rings (IR), respectively.  
}
\begin{tabular}{ccccc}
\hline 
  & $M=7$ & $M=8$  &  $M=10$  &  $M=20$  \\
\hline
    $d_{eff}^{NR}$ & 2.10$\pm$0.02 & 2.26$\pm$ 0.05& 2.38$\pm$0.01&3.06 $\pm$  0.06\\
    $d_{eff}^{IR}$  & 2.42$\pm$0.01 & 2.56$\pm$0.01& 2.72$\pm$0.02& 3.62$\pm$ 0.08\\
    $d_g^{NR}$ & 1.84$\pm$   0.05 & 1.98$\pm$0.05& 2.24$\pm$0.07
& 3.18$\pm$ 0.11 \\
    $d_g^{IR}$   & 2.12 $\pm$  0.03& 2.26$\pm$ 0.04& 2.52$\pm$    0.05& 3.48$\pm$    0.09\\
\hline
\end{tabular}
\end{table}

A subpeak in the CM-RDF of IRs is observed at a position that is near or slightly less than the radius of gyration of sub-rings, representing the probability of finding a sub-ring around a reference sub-ring in same molecule.  The subpeak for the system of NRs is absent.  The appearance of subpeak originates from the fact that two interlocked sub-rings cannot be completely separated due to the permanent mechanical bonds.  The subpeak of IR-molecules system with $M=7$ is  more pronounced than the contact peak.  This indicates intramolecular correlation dominates the interactions between sub-rings at that chain length.  With increasing chain length, the subpeak decreases and becomes broader.  Eventually, it is lower than the contact peak.  Intermolecular correlation plays a major role in the sub-rings interactions. 

\begin{figure}[!t]
\centering
 \includegraphics[width=\linewidth]{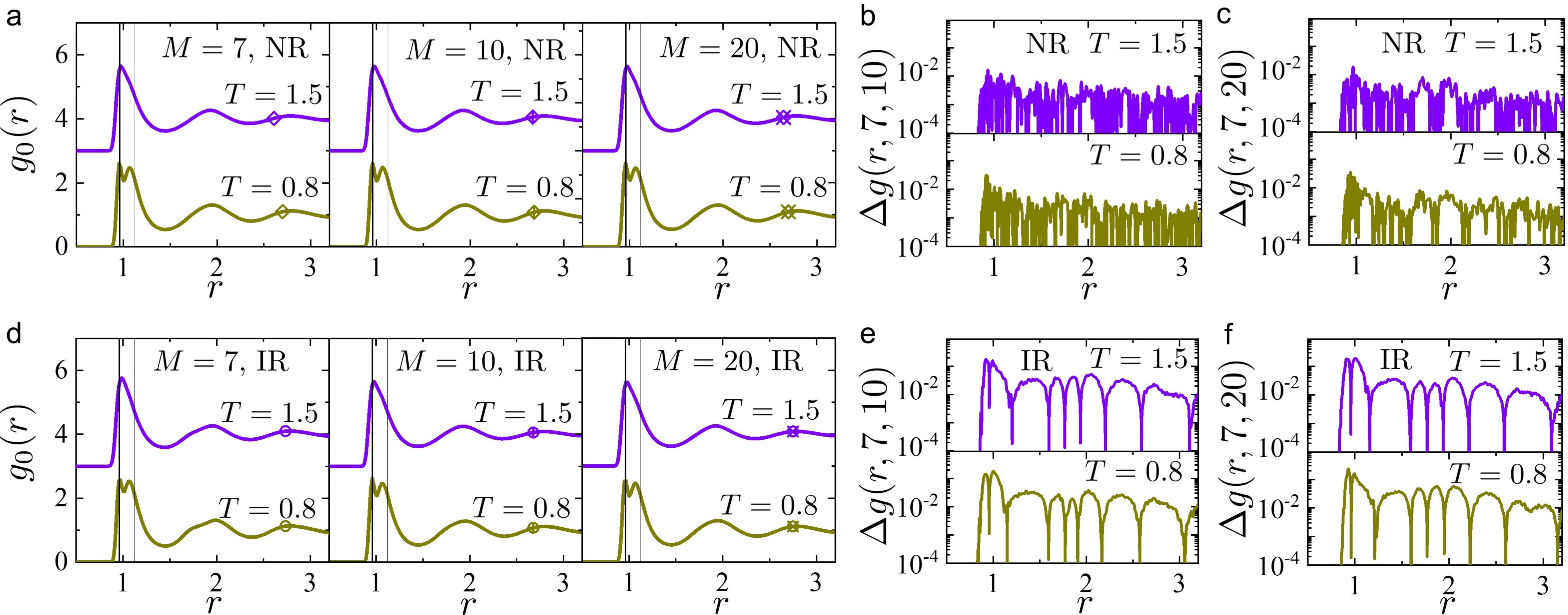}
	\caption{(a) Monomer-monomer radial distribution functions for nonconcatenated rings with $M=7$ (left), $M=10$ (middle) and $M=20$ (right) at two typical temperatures.  The simulation data for the system with $T=1.5$ are shifted up by 3 units for clarity. (b and c) The difference of MM-RDFs for the systems of nonconcatenated rings (NR) at $T=1.5$ (blue) and $T=0.8$ (red).  It is obtained by $\Delta g(r,M_1,M_2)=|g_0(r,M_1)-g_0(r,M_2)|$.  (d-f) The same as (a-c), but for the system of interlocked rings (IR).}
\label{fig:3}
\end{figure}

In general, local structures characterize the disordered packing of particles in a cage formed by their surrounding particles.  We study the local structures of monomers and the effect of mechanical bonds on the pair correlations by calculating the monomer-monomer RDF (MM-RDF), $g_0$.  In Figure \ref{fig:3}a and \ref{fig:3}d, the MM-RDFs show two split peaks around the monomer-monomer separation of $r=1$ both for the systems of interlocked and nonconcatenated rings at $T=0.8$.  This split results from two different length scales in the model polymer, bond length $r\approx 0.96$ and the cutoff of nonbonded monomer-monomer, $r\approx 1.12$.  The main peak of $g_0$ determines the transient number of nearest neighbors.  As increasing temperature to $T=1.5$, the second peak around $r=1.12$ decreases and the split is not visible, which is regarded as a sign of softening of a cage at high temperature.

We introduce a function characterizing the difference of the MM-RDFs of systems with different chain lengths, defined as $\Delta g(r,M_1,M_2)=|g_0(r,M_1)-g_0(r,M_2)|$, to explore how chain length affects the local structures of the system with mechanical bonds.   We report $\Delta g$ at fixed $M_1=7$ for $M_2=10$ and $M_2=20$, respectively.  The function $\Delta g^{NR}$ for the system of NRs is approximately $2\%$-$4\%$ for all particle separations studied (Figure \ref{fig:3}b and \ref{fig:3}c), indicating the MM-RDF is almost independent of chain length. While $\Delta g^{IR}$ for IRs reach values up to $17\%$ ($M_2=10$ ) and $20\%$ ($M_2=20$ ) at $r\approx 1$, this implies that the IR system of short chain has significantly difference on local monomer packing from long chain.  At higher temperature ,$T=1.5$, $\Delta g^{IR}$ remains almost unchanged.  We therefore believe the variation of MM-RDF on different chain lengths is mainly due to the combined effect of the temperature-independent chain length and permanent mechanically interlocking.

\subsection{Segmental motion}
\begin{figure}[!t]
\centering
	\includegraphics[width=0.5\linewidth]{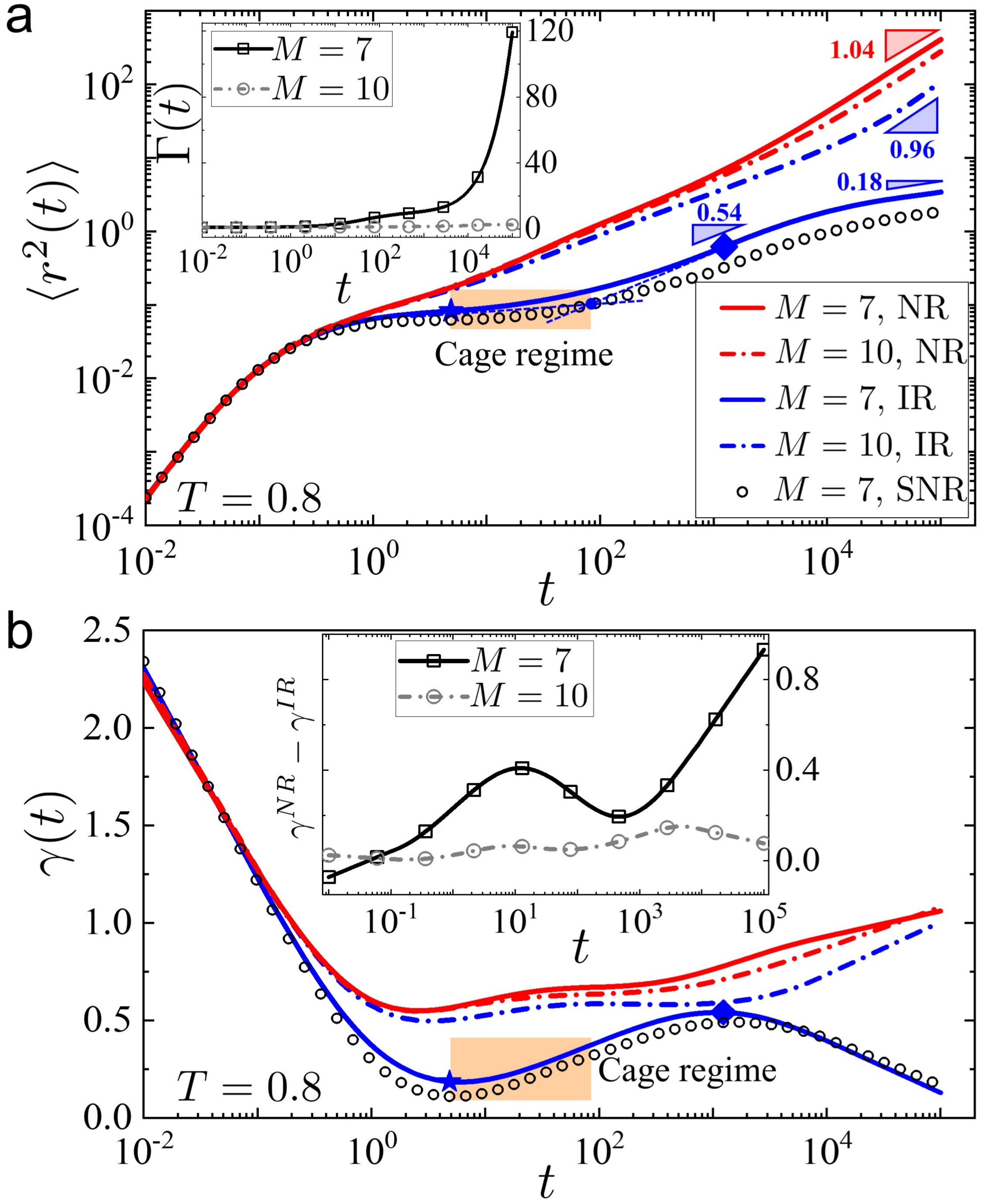}
	\caption{(a) Monomer mean squared displacements for the systems of nonconcatenated rings (NR, solid and dash-dot red lines) and interlocked rings (IR, solid and dash-dot blue lines).  Orange shaded area represents the cage regime.  The blue star and diamond illustrate the location of the minimum and maximum of the power exponents, respectively. The open circles represent the result of the system of semiflexible NRs (SNR).  Inset: the ratio of MSDs for the system of NRs to that of IRs, $\Gamma(t)=\langle r^2(t)\rangle^{NR}/\langle r^2(t)\rangle^{IR}$ for two typical chain lengths.  (b) The power exponents of MSDs as a function of time.  Corresponding cage regime is marked (orange shade).   Panel (b) shares the same legends shown in (a).  Inset: the difference of the power exponents for two typical chain lengths.}
\label{fig:4}
\end{figure}

We study segmental motion by investigating the mean square displacements (MSDs) of monomers
\begin{equation}\label{5}
    \langle r^2(t)\rangle=\frac{1}{M_{all}}\sum_{i=1}^{M_{all}}\langle [\bm{r}_i(t)-\bm{r}_i(0)]^2\rangle
\end{equation}
where $\bm{r}_i$ is the position of monomer $i$, $M_{all}$ is the total number of monomers in the studied system.  The angled bracket indicates an ensemble average.   

Figure \ref{fig:4}a shows MSDs for systems of nonconcatenated and interlocked rings with different chains at $T=0.8$, respectively.  At different time scales. the results of MSDs display three clearly different features.  First, at short time $t\ll 1$, all data sets are nearly identical and exhibit ballistic motion, $\langle r^2(t)\rangle\sim t^2$.  

Second, at intermediate times, the segmental motion appears to be subdiffusive following $\langle r^2(t)\rangle\sim t^{\gamma(t)}$.  For NRs, the time-dependent power exponents reach a minimum value of $\gamma\approx 0.55$ (Figure \ref{fig:4}b).  The behavior has been attributed to chain connectivity \cite{Chong2007,Frey2015}.  For the system of IRs with long chain ($M=10$), the amplitude of MSD has a slight reduction but the power exponents are similar with NRs.  In contrast, IRs with $M=7$ shows strong inhibition both in MSDs and the corresponding exponents.  A cage-like plateau emerges, corresponding to an apparent minimum in the predicted exponent, $\gamma\approx 0.18$ at $t\approx 5$ (star in Figure \ref{fig:4}b).  This minimum exponent marks the beginning of transient dynamic arrest.  Subsequently, the time-dependent exponent $\gamma(t)$ has a maximum at about $t\approx 1450$ (solid diamond in Figure \ref{fig:4}B), representing the escape of the interlocked monomers from the cage.  The end of cage regime can be defined by the intersection of two tangent lines at the two extremums in the log-log plot of MSD (solid circle in Figure \ref{fig:4}a).  We estimate the duration of caging regime, $t\approx4$-$85$, as indicated by the shade of orange in Figure \ref{fig:4}. 

At long times, the MSDs of NRs display a standard Fickian diffusion, where the power exponent approximately equals to $1$.  The interlocked one at $M=10$ also recovers to the same diffusion with $\gamma\approx 1$ but appears at a longer time.  The MSDs of the system of short IRs exhibit a strong suppression and corresponding power exponents have a dramatic reduction, $\gamma(t=10^5)\approx 0.18$, which are anomalous features specific to the system of mechanical bonds with rather short chain.  

We also find that the IR system shares the similar intermediate-time caging and long-time subdiffusive behavior to the semiflexible nonconcatenated rings (SNR) with real local stiffness as shown in Figure \ref{fig:4}.  In conventional polymer glasses,  chain stiffness results in a greater persistence length, that is, a larger number of dynamically correlated segments along chain backbone. \cite{Rubinstein2003} At the intermediate time ($\beta$ regime), a tagged segment is trapped in a cage formed by the surrounding particles.  Escaping the cage to move greater distance requires rearrangements of these neighbors, which are determined by the dynamically correlation length.  This explains why semiflexible chains with more correlated segments have stronger cage effect and is responsible for glassy behaviors, such as dynamic fragility and slowing dynamics \cite{Colmenero2008,Saltzman2007}. 

The effect of mechanical bonds on different chain lengths can be highlighted by introducing the MSDs ratio of NRs and IRs at the same chain length, $\Gamma(t)=\langle r^2(t)\rangle^{NR}/\langle r^2(t)\rangle^{IR}$.  The ratio for the system with short chain ($M=7$) increases to $\sim120$ at long times (inset of Figure \ref{fig:4}a), while that with $M=10$ remains almost constant and its value does not exceed $3$ throughout the studied time scale.  The corresponding change in scaling behavior can be quantified by the difference in the exponent, $\Delta \gamma=\gamma^{NR}(t)-\gamma^{IR}(t)$.  The exponent difference is small at $M=10$ and does not exceed $0.16$ (inset of Figure \ref{fig:4}b).  At $M=7$, $\Delta \gamma$ reaches two maximums at intermediate time and long time, $\Delta \gamma(t= 10)\approx 0.4$ and $\Delta \gamma(t= 10^5)\approx 0.9$, respectively.

\subsection{Motion of sub-rings}
\begin{figure}[!t]
\centering
	\includegraphics[width=0.5\linewidth]{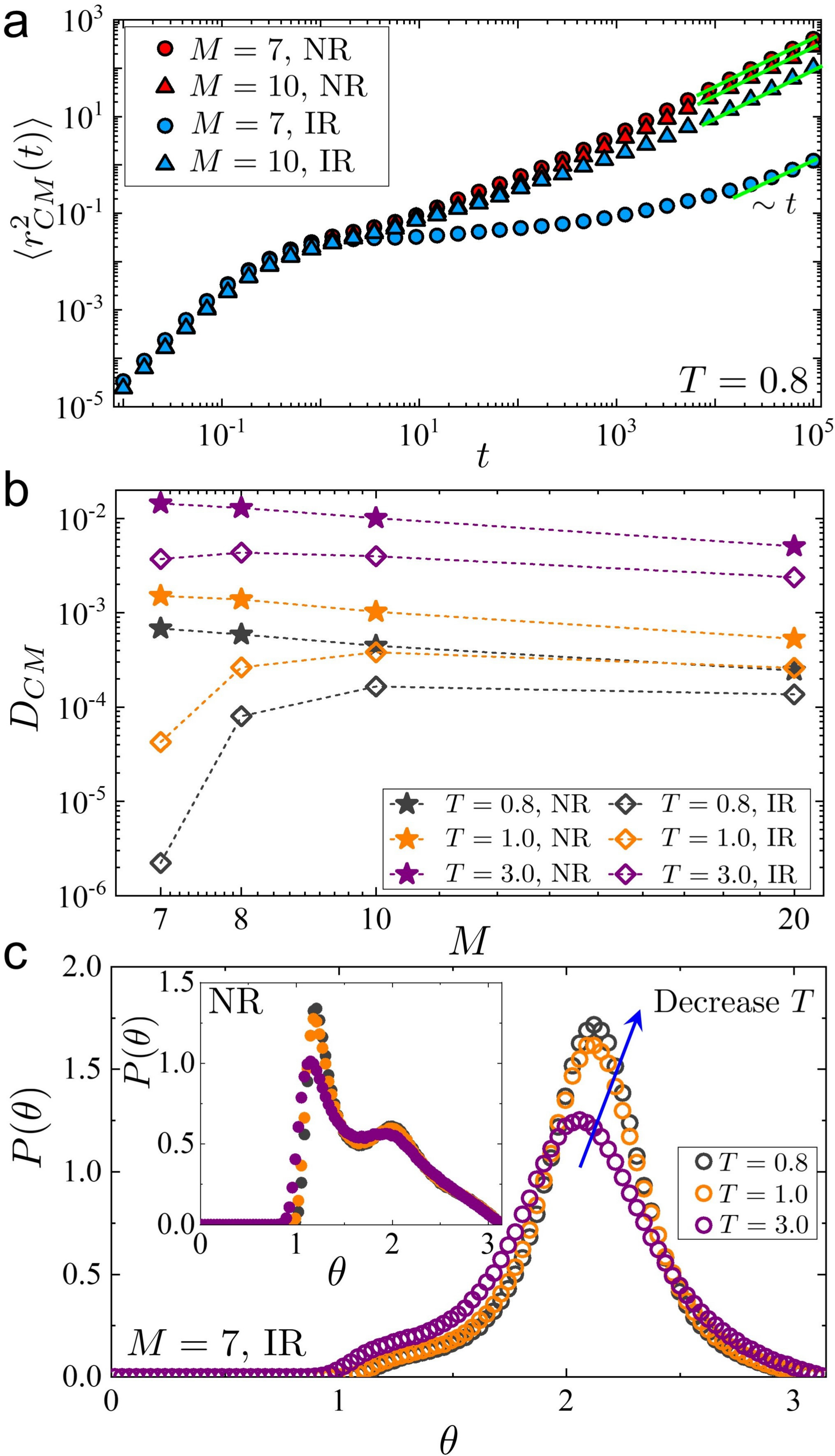}
	\caption{(a) The center-of-mass mean squared displacements of sub-rings for a few typical sub-ring lengths.  (b) The center-of-mass diffusion constants as a function of sub-ring length for a few typical temperatures. (c) The bond angle distributions for a few typical temperatures.}
\label{fig:5}
\end{figure}

The motion of sub-rings is observed by measuring the center-of-mass MSD (CM-MSD), $\langle r^2_{CM}(t)\rangle=\sum_{j=1}^{M_{c}}\langle [\bm{R}_j^{CM}(t)-\bm{R}_j^{CM}(0)]^2\rangle$/$M_c$, where $\bm{R}^{CM}_j$ is the position of the CM of the $j$th sub-ring, $M_c$ is the number of sub-rings in our simulation. The surrounding rigid IRs with $M=7$ leads to an apparent intermediate-time plateau in the CM-MSD as shown in Figure \ref{fig:5}a.  In contrast to subdiffusive behavior in MM-MSD, the scaling exponents in CM-MSD increase and approach to $1$ at long times, indicating that CM diffusion is Fickian.  We also notice that the amplitude in the CM-MSD is smaller than that of corresponding MM-MSD at long times.  These polymer-specific anomalies in MSD are suggested to arise from chain connectivity, reflected by conformational relaxation due to intrachain cooperative motion \cite{Rubinstein2003,Doi1988,Chong2007}.

For flexible and nonconcatenated rings, the long-time Fickian diffusion constant, $D_{CM}=lim_{t\to\infty}\langle r^2_{CM}(t)\rangle/6t$, monotonically decreases with chain length shown as the solid stars in Figure \ref{fig:5}b. In previous MD simulation for a melt of rings, the diffusion constant showed a power-law decay, $D_{CM}\sim M^{-1.34}$ \cite{Halverson2011Dynamics}. Interestingly, as chain length increases, $D_{CM}$ for the system of IRs shows nonmonotonic behavior (open diamonds in Figure \ref{fig:5}b). One possible reason for the anomalous diffusion is that the extent of inhibition of sub-rings motion is different in the systems with different chain lengths. Mechanical bonds strongly restrict the motion of interlocked sub-rings with $M=7$ and result in a sharp decrease of diffusion. With increasing chain length, mechanically interlocked constraints become weaker and the chain diffusion grows. For the system with longer chain, mechanical bonds have almost no effect and the diffusion constant of interlocked sub-rings is expected to decay with chain length.  Moreover, the nonmonotonic behaviors become less pronounced at high temperature ($T=3.0$), implying that chain length and temperature are two variables that have opposite effects on sub-ring motion.  Chain stiffness is reduced with temperature, manifested by the decrease of peak value of the bond angle distributions at high temperature (Figure \ref{fig:5}c).

\begin{figure}[!t]
\centering
	\includegraphics[width=\linewidth]{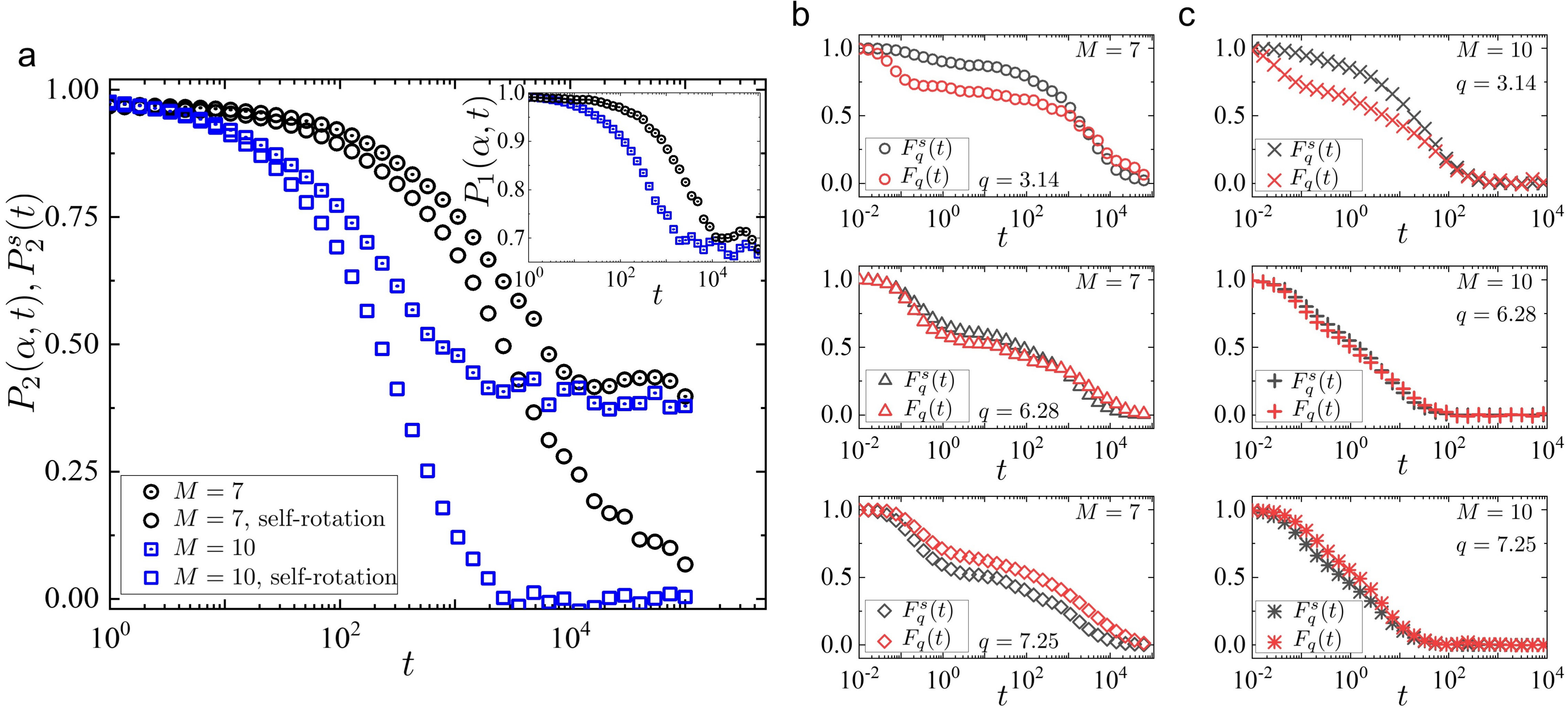}
	\caption{(a) Time correlations of the angle, $P_2$, and $P_1$ (inset), between the two sub-rings belonging to the same IR molecule and orientational autocorrelation functions of a sub-ring at $M=7$ and $M=10$.  (b and c) Incoherent functions and collective functions of the system of IRs with $M=7$ and $M=10$ at different wave vectors.}
\label{fig:6}
\end{figure}

We next invesitigate the rotation of sub-rings.  We first define the orientation of a ring by a vector $\bm{e}$ joining the monomer $1$ and $(M-1)/2$ ($M$ is odd) or $M/2$ ($M$ is even).  The two vectors, $\bm{e}_1$ and $\bm{e}_2$, represent the orientations of the two sub-rings in an IR molecule.  The mutual rotation of the two interlocked sub-rings can be quantified by the rotational correlation functions \cite{Moreno2005}, $P_1(\alpha,t)=\langle cos[\alpha(t)-\alpha(0)]\rangle$ and $P_2(\alpha,t)=\langle 3cos^2[\alpha(t)-\alpha(0)]-1\rangle/2$, where $\alpha$ is the relative angle between the above two vectors belonging to the same IR molecule.  The relaxation of the relative angle arises from the rotation of the sub-ring in its plane. As shown in Figure \ref{fig:6}a, the rotational correlations $P_1$ and $P_2$ for the IR systems are frozen at $\sim 0.67$ and $\sim 0.4$, respectively.  This indicates that the mutual rotation of sub-rings in the same IR molecule cannot occur.  Here, we provide a physical picture for the frozen rotation.  When a sub-ring rotates in its plane, its monomers should penetrate the void formed by another sub-ring in the same IR molecule.  For IR system with $M=7$, the void size in the middle of sub-ring is smaller than a monomer.  Thus, mechanical interlocking leads to the inhibition of the mutual rotation of short sub-rings.  Whereas for longer chain, though the void size is larger, the conformation of a sub-ring in the dense melts is collapsed, which suppresses the mutual rotation in an IR molecule. 

We characterize the self-rotation of a sub-ring by examining the following orientational autocorrelation function, $P_2^s(t)=\langle 3[\bm{e}_s(t)\cdot \bm{e}_s(0)]^2-1\rangle/2$, where $\bm{e}_s(t)$ is the orientation of the tagged sub-ring at time $t$.  $P_2^s(t)$ can decay to zero at long time as shown in Figure \ref{fig:6}a.  This reveals that overall rotational motion occurs in an IR molecule.  

To provide more insights to the motion of sub-rings, we check the coherent function, defined as 
\begin{equation}\label{6}
    F_q(t)=\frac{1}{S(q)M_{all}}\sum_{i,j=1}^{M_{all}}\Big\langle exp\big[i \bm{q}\cdot(\bm{r}_j(t)-\bm{r}_i(0))\big]\Big \rangle
\end{equation}
The function represents the collective relaxation of monomers on the corresponding length scale, $l=2\pi/|\bm{q}|$.  When $j$th monomer at time $t$ moves to where $i$th monomer locates at t=0, the collective scattering function cannot decay to zero.  Therefore, the mutual rotation is espected to cause the freezing of $F_q$ at long time.  We also compute the incoherent function by the following expression,
\begin{equation}\label{6}
    F_q^s(t)=\frac{1}{M_{all}}\sum_{i=1}^{M_{all}}\Big\langle exp\big[i \bm{q}\cdot(\bm{r}_i(t)-\bm{r}_i(0))\big]\Big \rangle
\end{equation}
Figure \ref{fig:6}b and \ref{fig:6}c show that the intermediate-time plateaus of these functions are more apparent at small wavevectors.  At longest time we studied, the incoherent and coherent functions can decay to zero and their coulpings are observed at different wavevectors.  This further supports the finding about the inhibition of the mutual rotation in an IR molecule.  

\subsection{Segmental Relaxation and Localization}
\begin{figure}[!t]
\centering
	\includegraphics[width=\linewidth]{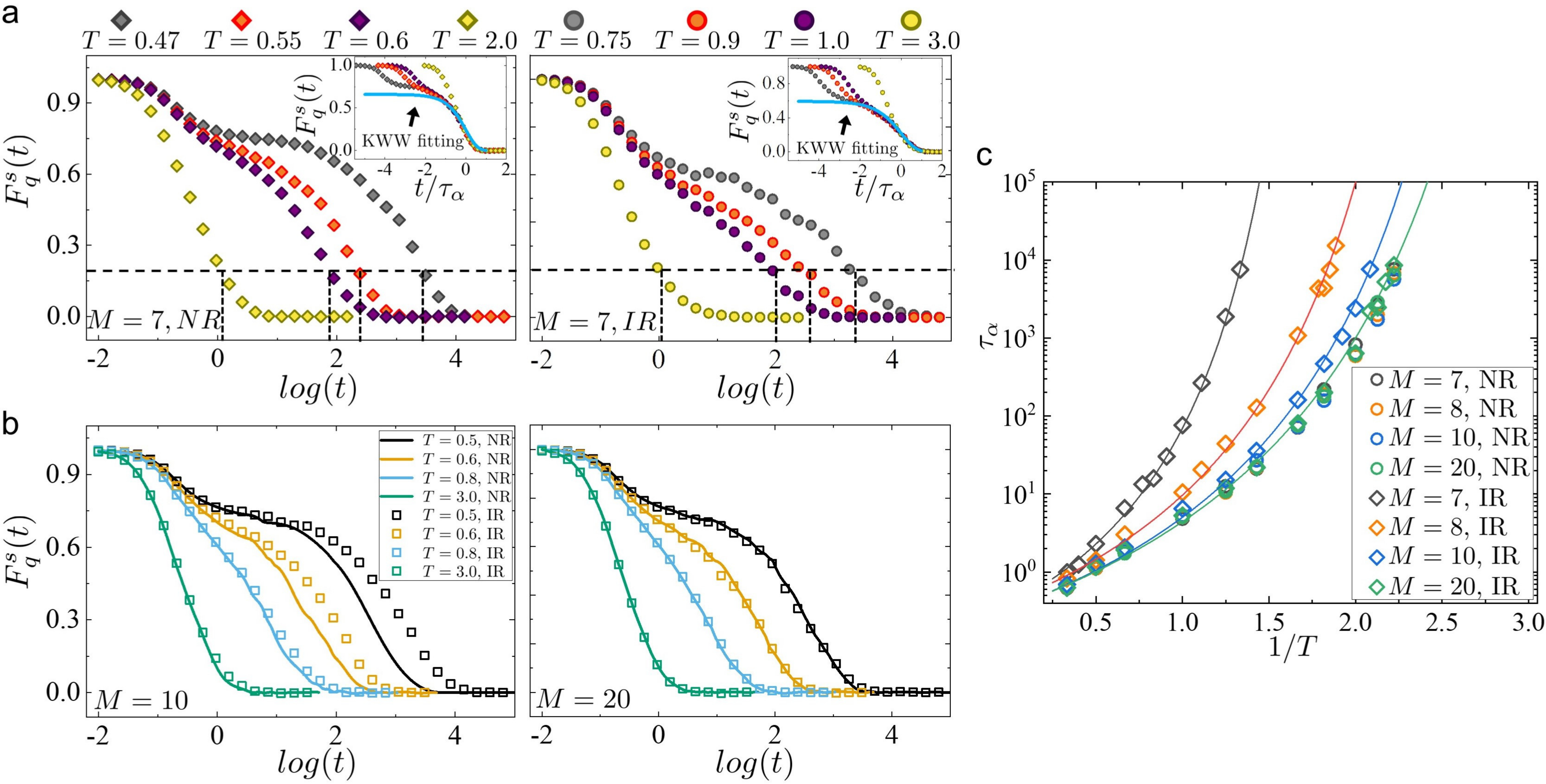}
	\caption{(a) Left: incoherent functions for the system of nonconcatenated rings with $M=7$ at a few typical temperatures. Inset:  incoherent functions as a function of reduced time $t/\tau_{\alpha}$.   The thick blue line is the KWW fit of eq \ref{7}.  Right: the same as in the left panel, but for the interlocked rings system.  The left and right panels share the same vertical coordinate axis. (b) Incoherent functions of NRs (solid lines) and IRs (open squares) at $M=10$ (left) and $M=20$ (right), respectively. The left and right panels share the same legend. (c) Segmental relaxation time as a function of the reciprocal of temperature $1/T$.  The solid lines represent the VFT fitting for the system of interlocked rings.}
\label{fig:7}
\end{figure}

There is coulping in the coherent-incoherent functions, we therefore  use only the incoherent function to characterize the relaxation dynamics on segmental scale by taking wavevector as $|\bm{q}|=2\pi/\sigma=6.28$.  At high temperature, $F_q^s(t)$ typically decays as $\sim exp(-t/\tau_{\alpha})$, which is a universal feature of normal liquids.  With decreasing temperature, we observe an intermediate time plateau in the incoherent function (Figure \ref{fig:7}a and \ref{fig:7}b), resulting from transient localization.  The plateau is more apparent for lower temperature.  The relevant $\alpha$-relaxation time can be empirically defined by $F_q^s(t=\tau_{\alpha})=0.2$ shown as the dashed lines of Figure \ref{fig:7}a.   $F_q^s(t)$ are plotted at four representative temperatures in Figure \ref{fig:7}a.  The interlocked rings require higher temperature in comparison to the nonconcatenated ones to achieve similar $\alpha$-relaxation times.  The most striking difference in $F_q^s(t)$ is that the system of IRs has a less apparent intermediate time plateau, but their long-time relaxations remain the same value.  

To further describe the effect of mechanical bonds on the incoherent functions, we apply the empirical Kohlrausch-Williams-Watts (KWW) formula \cite{Kob1994,Binder1998}
\begin{equation}\label{7}
    F_q^s(t)=Aexp\Bigg[-\bigg(\frac{t}{\tau_{\alpha}}\bigg)^{-\beta}\Bigg]
\end{equation}
where $A$ is amplitude, representing the height of intermediate plateau.  $\beta$ is the KWW exponent and lies in the range,  $0<\beta<1$.  The more the exponent deviates from $1$, the more the long-time relaxation becomes stretched.  The plots of $F_q^s$ versus $t/\tau_{\alpha}$ for different temperatures collapse onto a master curve at long times (insets in Figure \ref{fig:7}a), which is well known as time-temperature superposition \cite{Kob1994}.  The parameters $A$ and $\beta$ are independent of temperature at long time.  We obtain $A=0.659$ and $\beta=0.756$ for the system of NRs at $M=7$, while $A=0.592$ and $\beta=0.589$ for that of interlocked ones.  Therefore, the system of IRs has a lower plateau in the incoherent function and more stretched relaxation, which is consistent with those found in linear chain with intramolecular barriers \cite{Colmenero2008}. 

In Figure \ref{fig:7}b, we compare segmental relaxations for systems of NRs (solid lines) and IRs (open squares) at chain lengths of $M=10$ and $M=20$, respectively.  NRs with $M=10$ have faster segmental relaxations than interlocked ones when $T\le 0.8$ (left panel in Figure \ref{fig:7}b).  For $M=20$, $F_q^s(t)$ of the above two systems coincide at all temperatures studied (right panel in Figure \ref{fig:7}b).  This illustrates that mechanical bonds have little effect on segmental dynamics for long chain.  

We turn to the temperature dependence of $\alpha$-relaxation time.  As shown in Figure \ref{fig:7}c, at the same temperature, the system of IRs with short chain ($M\leq 10$) has larger relaxation time than that of NRs.  With increasing chain length to $M=20$, the curves of $\tau_\alpha(T)$ of IRs (open diamonds in Figure \ref{fig:7}c) approach to that of nonconcatenated ones (open circles in Figure \ref{fig:7}c),  which have almost identical temperature dependence on relaxation time for different chain lengths. 

In general, the exponential increase in $\tau_{\alpha}$ with cooling can be quantified by the empirical Vogel–Fulcher–Tammann (VFT) relation,
\begin{equation}\label{8}
    \tau_{\alpha}=\tau_0exp\Bigg[\frac{1}{B(T/T_0-1)}\Bigg]
\end{equation}
where $\tau_0$, $T_0$ and $B$ are free parameters.  The VFT relation agrees well with our simulation results for the system of IRs (solid lines in Figure \ref{fig:7}c).  The relaxation time diverges at the glass transition temperature $T_0$, marking ergodicity breaking.  The change in $T_0$ for the system of NRs with chain length is modest, $T_0^{NR}=0.261$-$0.267$ as listed in Table \ref{tab:2}.  For IRs system, the transition temperature has a higher value $\sim0.471$ at $M=7$, and reduces to a value similar to that of nonconcatenated ones, $\sim0.265$ at $M=20$. 

The fragility index $B$ in VFT relation quantifies the growth rate of relaxation time with decreasing temperature.  Higher $B$ corresponds to more fragile glass, that is, a greater deviation from the Arrhenius behaviors \cite{Sastry2001}.  Compared to the nonconcatenated rings, mechanical bonds induce more fragile glass-forming rings.  For example, the values of fragility index are about $B=0.143$-$0.149$ in the system of NRs, whereas $B$ reaches $0.171$ for the system of IRs with $M=7$ (Table \ref{tab:2}).  

\begin{table}[!t]
\caption{\label{tab:2}%
Summary of VFT glass transition temperature, $T_0$, fragility index, $B$, and the critical Lindemann-like oscillation distance, $u_0$, for the systems of nonconcatenated rings (NR) and interlocked rings (IR), respectively.}
\begin{tabular}{ccccc}
 \hline 
    & $M=7$  &  $M=8$  &  $M=10$ & $M=20$\\
 \hline
    $T_0^{NR}$ & 0.261 & 0.267 & 0.262 & 0.263 \\
    $T_0^{IR}$ & 0.471 & 0.329 & 0.289 & 0.265\\
    $B^{NR}$ & 0.144 & 0.147& 0.149&0.143 \\
    $B^{IR}$ & 0.171 & 0.155 & 0.152& 0.142\\
    $u_0^{NR}$ & 0.406 & 0.402& 0.403& 0.405\\
    $u_0^{IR}$ & 0.537 & 0.440& 0.411& 0.403\\
 \hline
\end{tabular}
\end{table}

The localization model (LM) provides a quantitative approach to describe the long-time $\alpha$-relaxation in terms of the free-volume perspective, that is, short-time local rattling motion within a cage.  The rattling motion is quantified by the Debye-Waller factor $\langle u^2\rangle$.  Within the LM, segmental relaxation relates to Debye-Waller factor by following expression \cite{Douglas2015PNAS}, 
\begin{equation}\label{9}
    \tau_{\alpha}=\tau_\eta exp\Bigg[\bigg(\frac{u_0^2}{\langle u^2\rangle}\bigg)^\eta\Bigg]
\end{equation}
where $\tau_\eta$ is an intrinsic time constant.  $u_0$ has a unit of length and is regarded as a critical Lindemann-like oscillation distance.  The exponent $\eta$ characterizes the free-volume anisotropy.  Under the isotropic spherical free-volume assumption, one would expect $\eta=3/2$ \cite{Douglas2021}.  In our calculation, the Debye-Waller factor is extracted from MSDs, $\langle u^2\rangle=\langle r^2(t=1)\rangle$, according to recent work \cite{Douglas2015PNAS,Douglas2021}. 

\begin{figure}[!t]
\centering
	\includegraphics[width=0.9\linewidth]{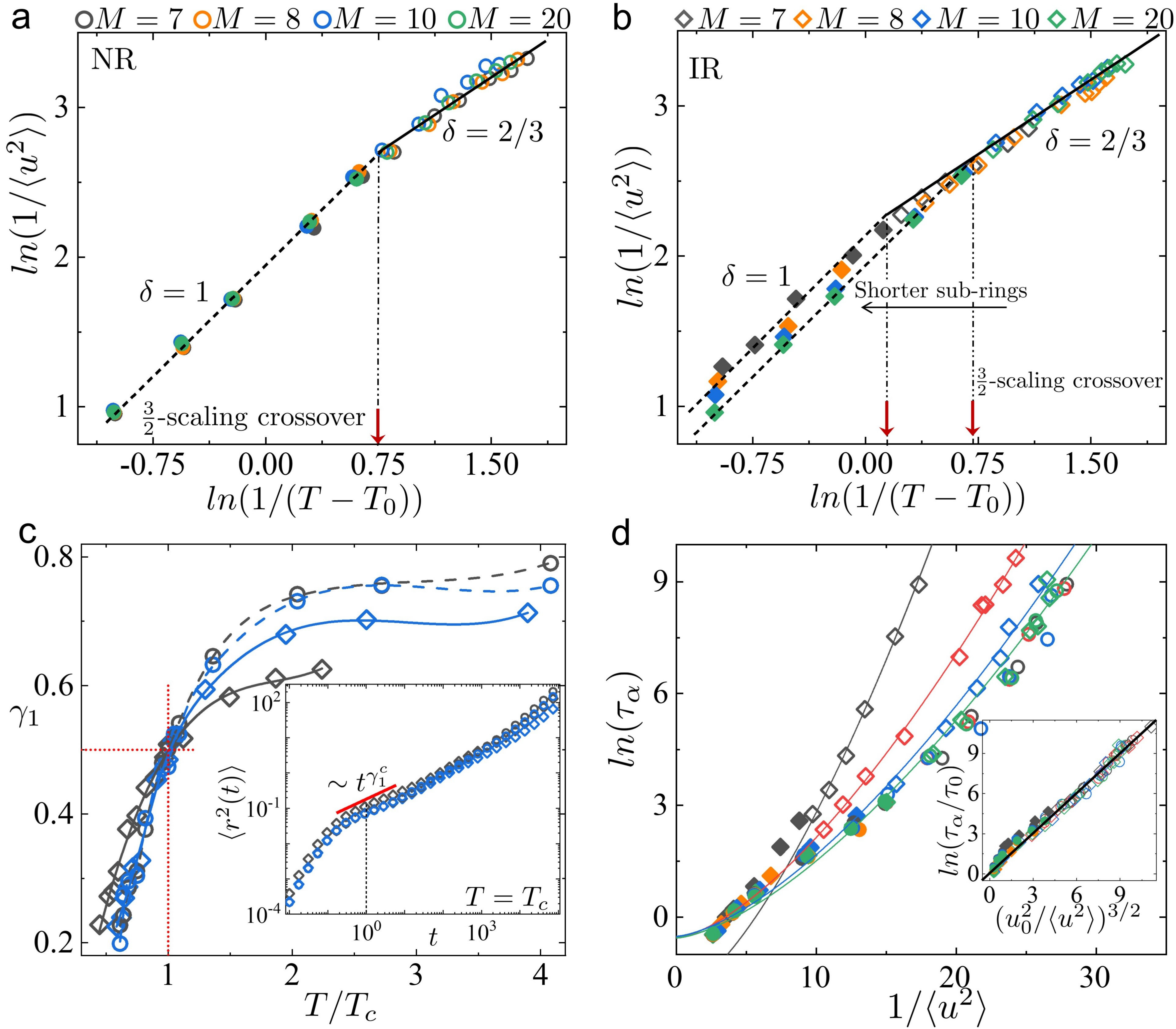}
	\caption{(a) The log-log plot between the reciprocal of the Debye-Waller factor and the glassy depth, $\xi=1/(T-T_0)$, for nonconcatenated rings (NR).  There are two distinct scaling regimes with different power exponents, $\delta=1$ (solid symbols) and $\delta=2/3$ (open symbols).  Dashed lines and solid lines are guide for eyes.  The vertical dash-dot line and red arrow indicate the location of scaling crossover.  (b) The same as (a), but for interlocked rings (IR).  The location of scaling crossover is shifted to smaller glassy depth for shorter sub-rings. (c) The scaling exponents of MSD at  $t=1$ as a function of reduced temperature $T/T_c$.  The panel shares the same legends shown in (a) and (b).  The solid lines and dashed lines are only used to guide to the eye.  The cross point of the horizonal and vertical red-dotted lines indicates the critical scaling exponents $\gamma_1^c=0.5$ at the $3/2$-scaling crossover temperature. Inset: MSD for the above four systems at $T=T_c$.  The thick red line represents the critical slope of MSD at $t=1$ and indicates the occurring of caging. (d) Relaxation time as a function of the reciprocal of the Debye-Waller factor.  The solid lines represent the $\eta=3/2$ fitting curves in eq \ref{9}.  Inset: collapsed data for the scaled relaxation time versus $(u_0^2/\langle u^2\rangle)^{3/2}$.  The black solid line represents their linear relationship.  The solid symbols represent relaxations in $\eta=1$ scaling regime. }
\label{fig:8}
\end{figure}

We first study the temperature dependence of $\langle u^2\rangle$.  Combining eq \ref{8} and eq \ref{9}, we directly obtain the following scaling relation between localization and temperature, $1/\langle u^2\rangle\sim (1/(T-T_0))^{\delta}$, where $\delta=1/\eta$.  To test the validity of the above relation, we plot our simulation results for the reciprocal of Debye-Waller factor, $1/\langle u^2\rangle$, as a function of glassy depth $\xi=1/(T-T_0)$, denoting how close the temperature is to $T_0$.  The slope in the log-log plot indicates the scaling exponent $\delta$ (Figure \ref{fig:8}a and \ref{fig:8}b).  Surprisingly, there are two distinct regimes with different scaling exponents $\delta$.  At high temperature (small $\xi$), the localization-temperature scaling exponent is found to be $\delta=1$.  When temperature decreases towards $T_0$ (large $\xi$), the ring polymer systems have a different scaling behavior, $\delta=2/3$, i.e., $\eta=3/2$, shown as the solid lines in Figure \ref{fig:8}a and \ref{fig:8}b.  Here, we consider the observed scaling crossover can be regarded as a signature, which system undergoes a high- to low-temperature transition in distinct dynamic regimes, from normal liquids to supercooled liquids.  

For the system of nonconcatenated rings, Figure \ref{fig:8}a shows that the scaling crossover occurs nearly at a constant of glassy depth, $\xi_c^{NR}\approx 2.11$ ($T_c^{NR}\approx0.735$), for all chain lengths studied.  While for the system of interlocked rings with short chain ($M<20$), we find that 
mechanical bonds reduce the crossover glassy depth to lower values (higher temperature) relative to that of nonconcatenated rings systems.  For example, $\xi_c^{IR}\approx 1.14$ ($T_c^{IR}=1.34$) at $M=7$, revealing that the mechanical bonds expand the $3/2$ $\langle u^2\rangle$-$\xi$ scaling region.  This corresponds to a broader range of supercooled liquids.  

To understand what is happening at the crossover temperature $T_c$, we investigate the scaling exponent, $\gamma_1$, of MM-MSD at the position of $\langle u^2\rangle$ at different temperatures.  The scaling exponents can be applied to indicate the emergence of caging dynamics.  Figure \ref{fig:8}c shows that the scaling exponent increases dramatically as temperature increases to $T_c$ and almost saturates at $T/T_c>2$.   Interestingly, we find the scaling exponent remains constant $\gamma_1^c\approx 0.5$ at the crossover temperature whether interlocked or not, reflecting there are similar caging dynamics around $T_c$.  We further suggest that the crossover temperature reveals the transition of caging dynamics, from weak to strong blocking.  Therefore, we believe the localization model with $\eta=3/2$ is generally applicable at strong local cage below $T_c$. 

In the following, we turn to examine the validity of LM approach in eq \ref{9} under the  isotropic free-volume assumption ($\eta=3/2$).  Figure \ref{fig:8}c shows the $\alpha$-relaxation time as a function of reciprocal of Debye-Waller factor $1/\langle u^2\rangle$ for nonconcatenated and interlocked ring systems, respectively.  The solid lines indicate the fits to eq \ref{9} for IRs when $\eta=3/2$.  The fitting curves agree well with the relaxation data in $3/2$-scaling regime (open diamonds).  For the same Debye-Waller factor, rings with mechanical bonds show higher relaxation times.  When plotting the scaled relaxation time, $\tau_{\alpha}/\tau_0$, as a function of $(u_0^2/\langle u^2\rangle)^{3/2}$, all data collapse onto a master curve, except for those points in the normal liquid regime, which lie above the straight line predicted by eq \ref{9} (inset of Figure \ref{fig:8}c).  The localization model with isotropic free-volume assumption describes the relaxation data in deeply supercooled regime  well.

The corresponding critical oscillation distance $u_0$ are listed in Table \ref{tab:2}.
With approaching to the glass transition $T_0$, the amplitude of local monomer vibration decreases to $u_0$, marking that the system undergoes a transition from delocalized to localized motion.  We show $u_0^{NR}\approx0.402$-$0.406$ for NRs.  The system of IRs  with short chain has a higher critical length, $u_0^{IR}\approx 0.537$ at $M=7$.  This reveals that mechanical bonds can induce localized state at larger monomer vibration motion that makes it easier to enter supercooled regime. 

Finally, let us discuss the role of chain stiffness induced by mechanical interlocking on glass transition.  In fact, the effect of chain stiffness induced by real intramolecular barriers on polymer glass formation has been widely studied \cite{Kumar2013,Hong2015,Sun2018,Moreno2009,Moreno2011,Colmenero2008,Colmenero2015,Xu2020,Kenneth2004,Saltzman2007}.  Although a fundamental understanding of the role of chain stiffness on glass transition remains distinct in different theoretical perspectives,  the qualitative effects on glassy features are generally recognized, such as the increase of characteristic glass temperatures, localization lengths, relaxation times and dynamic fragilities.  These results are compatible with our findings in MIRs with stiffness induced by mechanical interlocking.  

Combined with Adam-Gibbs scenario and the string model \cite{Starr2014}, structural relaxation time is related to activation free energy $\Delta G$ and characteristic reduced string length $L$ through following expression, $\tau_{\alpha}=\tau_0exp(\Delta G L/k_BT)$.  By molecular dynamic simulations, Xu \emph{et al}. found that the string length $L$ significantly increases with  chain stiffness, thereby leading to the increase of relaxation time and fragility \cite{Xu2020}.   Schweizer and Saltzman developed nonlinear Langevin equation theory (NLET) for polymer glasses with chain stiffness \cite{Kenneth2004,Saltzman2007}.  In the framework of NLET, chain stiffness enters the activated barriers through dynamically correlated parameter $a_c$. Relaxation time displays significant increase according to the formula, $\tau_{\alpha}=\tau_0exp(a_c F_B/k_BT)$.  In mode-coupling theory (MCT), temperature, density and chain stiffness are considered as control parameters of a sharp ergodic-nonergodic transition. \cite{Colmenero2015,Colmenero2008} MCT predicts that relaxation time follows a power law around MCT transition temperature, 
$\tau_{\alpha}=(T-T_{MCT})^{-\gamma_{MCT}}$.  In the polymer melts with intramolecular barrier, the competition of local packing and chain stiffness are considered to result in large MCT exponent $\gamma_{MCT}$ and the increase of localization length, MCT transition temperature and relaxation time \cite{Colmenero2008}.  

\section{Conclusion}
In summary, we have investigated static structures and segmental dynamics influenced by the chain stiffness induced by mechanical interlocking in MIPs.  Our model system of the permanent interlocked rings exhibits unique glassy behaviors in rather short chain, which is very different from the existing chain-scale topological glass transition where transient interpenetration plays important role and requires the limit of long rings.  We demonstrate that the combined effect of mechanical bonds and chain length is to induce an effective local stiffness on the sub-rings.  The induced stiffness provides a physical explanation to the intermediate-length bump in intra-ring correlations, transient dynamic arrest on segments and the long-time suppression of MSD scaling exponents.  In addition, we also find that the IR system with $M=7$ exhibits significant difference on local monomer packings and the slowing of sub-rings motion, leading to nonmonotonic diffusivity with chain length.  For the flexible IR molecules, the freezing of relative orientational correlations and coupling of incoherent-coherent density correlation reveal the inhibition of mutual rotation of the two sub-rings in a molecule.

According to the VFT relation, the system of IRs has higher glass transition temperature and becomes more fragile.  Combined VFT equation and the localization model, a scaling relation between localization and temperature is observed.  Our simulation results have revealed a universal scaling crossover from normal liquid with exponent $\eta=1$ to supercooled liquid with $\eta=3/2$ when glassy depth increases.  At the crossover glassy depth, we find the critical scaling exponents are approximately equal to $0.5$ whether interlocked or not, suggesting that there is a transition from weak to strong caging.  Based on the free-volume perspective in the localization model, relating local rattling motion in a cage to long-time segmental dynamics, we have rationalized these findings about the effect of mechanical bonds on structural relaxation.  The system with mechanical bonds is supposed to vitrify more easily by lowering the crossover glassy depth and increasing the critical oscillation distance.  Induced chain stiffness helps us better understand the variations of the above glass properties. 

There are a few issues are worth exploring in the future. First, a microscopic theory including induced chain stiffness needs to be developed to deal with vitrification in mechanically interlocked polymer rings with varying topological entanglements \cite{Sussman2011}.  Second, further study on the similarities and differences between induced stiffness and real stiffness on glass transition in ring polymers is called for.  Third, are the scaling crossover in localization-temperature widespread among various glassy systems?, what physical quantity does the scaling exponent depend on at the microscopic level?.  Finally, mechanical interlocking can be regarded as a means of inducing local stiffness on monomers.  It is  helpful for the design of topologically-driven glassy materials with controllable static features, slow dynamics and rheological properties in the future. \cite{Deng2022,Zhang2022NC,Hagita2022}

\begin{acknowledgement}

J. L. acknowledges the support by National Natural Science Foundation of China no. 11804085.  B. Z. acknowledges the supports by National Natural Science Foundation of China nos. 11904320 and 11847115.  

\end{acknowledgement}

\bibliography{ACS_Ma}

\providecommand{\latin}[1]{#1}
\makeatletter
\providecommand{\doi}
  {\begingroup\let\do\@makeother\dospecials
  \catcode`\{=1 \catcode`\}=2 \doi@aux}
\providecommand{\doi@aux}[1]{\endgroup\texttt{#1}}
\makeatother
\providecommand*\mcitethebibliography{\thebibliography}
\csname @ifundefined\endcsname{endmcitethebibliography}
  {\let\endmcitethebibliography\endthebibliography}{}
\begin{mcitethebibliography}{74}
\providecommand*\natexlab[1]{#1}
\providecommand*\mciteSetBstSublistMode[1]{}
\providecommand*\mciteSetBstMaxWidthForm[2]{}
\providecommand*\mciteBstWouldAddEndPuncttrue
  {\def\EndOfBibitem{\unskip.}}
\providecommand*\mciteBstWouldAddEndPunctfalse
  {\let\EndOfBibitem\relax}
\providecommand*\mciteSetBstMidEndSepPunct[3]{}
\providecommand*\mciteSetBstSublistLabelBeginEnd[3]{}
\providecommand*\EndOfBibitem{}
\mciteSetBstSublistMode{f}
\mciteSetBstMaxWidthForm{subitem}{(\alph{mcitesubitemcount})}
\mciteSetBstSublistLabelBeginEnd
  {\mcitemaxwidthsubitemform\space}
  {\relax}
  {\relax}

\bibitem[Rosa \latin{et~al.}(2019)Rosa, Di~Stefano, and Micheletti]{Rosa2019}
Rosa,~A.; Di~Stefano,~M.; Micheletti,~C. Topological Constraints in Eukaryotic
  Genomes and How They Can Be Exploited to Improve Spatial Models of
  Chromosomes. \emph{Front. Mol. Biosci.} \textbf{2019}, \emph{6}\relax
\mciteBstWouldAddEndPuncttrue
\mciteSetBstMidEndSepPunct{\mcitedefaultmidpunct}
{\mcitedefaultendpunct}{\mcitedefaultseppunct}\relax
\EndOfBibitem
\bibitem[Rosa and Everaers(2008)Rosa, and Everaers]{Rosa2008}
Rosa,~A.; Everaers,~R. Structure and dynamics of interphase chromosomes.
  \emph{PLoS. Comput. Biol.} \textbf{2008}, \emph{4}, e1000153\relax
\mciteBstWouldAddEndPuncttrue
\mciteSetBstMidEndSepPunct{\mcitedefaultmidpunct}
{\mcitedefaultendpunct}{\mcitedefaultseppunct}\relax
\EndOfBibitem
\bibitem[Tezuka(2013)]{Tezuka2013}
Tezuka,~Y. \emph{Topological Polymer Chemistry}; World Scientific, 2013\relax
\mciteBstWouldAddEndPuncttrue
\mciteSetBstMidEndSepPunct{\mcitedefaultmidpunct}
{\mcitedefaultendpunct}{\mcitedefaultseppunct}\relax
\EndOfBibitem
\bibitem[Doi and Edwards(1988)Doi, and Edwards]{Doi1988}
Doi,~M.; Edwards,~S.~F. \emph{The Theory of Polymer Dynamics}; Oxford
  University Press, 1988\relax
\mciteBstWouldAddEndPuncttrue
\mciteSetBstMidEndSepPunct{\mcitedefaultmidpunct}
{\mcitedefaultendpunct}{\mcitedefaultseppunct}\relax
\EndOfBibitem
\bibitem[Rubinstein and Helfand(1985)Rubinstein, and Helfand]{Rubinstein1985}
Rubinstein,~M.; Helfand,~E. Statistics of the entanglement of polymers:
  Concentration effects. \emph{J. Chem. Phys.} \textbf{1985}, \emph{82},
  2477--2483\relax
\mciteBstWouldAddEndPuncttrue
\mciteSetBstMidEndSepPunct{\mcitedefaultmidpunct}
{\mcitedefaultendpunct}{\mcitedefaultseppunct}\relax
\EndOfBibitem
\bibitem[Zhou and Larson(2005)Zhou, and Larson]{Larson2005}
Zhou,~Q.; Larson,~R.~G. Primitive Path Identification and Statistics in
  Molecular Dynamics Simulations of Entangled Polymer Melts.
  \emph{Macromolecules} \textbf{2005}, \emph{38}, 5761--5765\relax
\mciteBstWouldAddEndPuncttrue
\mciteSetBstMidEndSepPunct{\mcitedefaultmidpunct}
{\mcitedefaultendpunct}{\mcitedefaultseppunct}\relax
\EndOfBibitem
\bibitem[Everaers \latin{et~al.}(2004)Everaers, Sukumaran, Grest, Svaneborg,
  Sivasubramanian, and Kremer]{Ralf2004}
Everaers,~R.; Sukumaran,~S.~K.; Grest,~G.~S.; Svaneborg,~C.;
  Sivasubramanian,~A.; Kremer,~K. Rheology and Microscopic Topology of
  Entangled Polymeric Liquids. \emph{Science} \textbf{2004}, \emph{303},
  823--826\relax
\mciteBstWouldAddEndPuncttrue
\mciteSetBstMidEndSepPunct{\mcitedefaultmidpunct}
{\mcitedefaultendpunct}{\mcitedefaultseppunct}\relax
\EndOfBibitem
\bibitem[Tzoumanekas and Theodorou(2006)Tzoumanekas, and
  Theodorou]{Tzoumanekas2006}
Tzoumanekas,~C.; Theodorou,~D.~N. Topological Analysis of Linear Polymer Melts:
  A Statistical Approach. \emph{Macromolecules} \textbf{2006}, \emph{39},
  4592--4604\relax
\mciteBstWouldAddEndPuncttrue
\mciteSetBstMidEndSepPunct{\mcitedefaultmidpunct}
{\mcitedefaultendpunct}{\mcitedefaultseppunct}\relax
\EndOfBibitem
\bibitem[Kapnistos \latin{et~al.}(2008)Kapnistos, Lang, Vlassopoulos,
  Pyckhout-Hintzen, Richter, Cho, Chang, and Rubinstein]{Kapnistos2008}
Kapnistos,~M.; Lang,~M.; Vlassopoulos,~D.; Pyckhout-Hintzen,~W.; Richter,~D.;
  Cho,~D.; Chang,~T.; Rubinstein,~M. Unexpected power-law stress relaxation of
  entangled ring polymers. \emph{Nat. Mat.} \textbf{2008}, \emph{7},
  997--1002\relax
\mciteBstWouldAddEndPuncttrue
\mciteSetBstMidEndSepPunct{\mcitedefaultmidpunct}
{\mcitedefaultendpunct}{\mcitedefaultseppunct}\relax
\EndOfBibitem
\bibitem[Halverson \latin{et~al.}(2011)Halverson, Lee, Grest, Grosberg, and
  Kremer]{Halverson2011Static}
Halverson,~J.~D.; Lee,~W.~B.; Grest,~G.~S.; Grosberg,~A.~Y.; Kremer,~K.
  Molecular dynamics simulation study of nonconcatenated ring polymers in a
  melt. I. Statics. \emph{J. Chem. Phys.} \textbf{2011}, \emph{134},
  204904\relax
\mciteBstWouldAddEndPuncttrue
\mciteSetBstMidEndSepPunct{\mcitedefaultmidpunct}
{\mcitedefaultendpunct}{\mcitedefaultseppunct}\relax
\EndOfBibitem
\bibitem[Halverson \latin{et~al.}(2011)Halverson, Lee, Grest, Grosberg, and
  Kremer]{Halverson2011Dynamics}
Halverson,~J.~D.; Lee,~W.~B.; Grest,~G.~S.; Grosberg,~A.~Y.; Kremer,~K.
  Molecular dynamics simulation study of nonconcatenated ring polymers in a
  melt. II. Dynamics. \emph{J. Chem. Phys.} \textbf{2011}, \emph{134},
  204905\relax
\mciteBstWouldAddEndPuncttrue
\mciteSetBstMidEndSepPunct{\mcitedefaultmidpunct}
{\mcitedefaultendpunct}{\mcitedefaultseppunct}\relax
\EndOfBibitem
\bibitem[Gómez \latin{et~al.}(2020)Gómez, García, and
  Pöschel]{Leopoldo2020}
Gómez,~L.~R.; García,~N.~A.; Pöschel,~T. Packing structure of semiflexible
  rings. \emph{Proc. Natl. Acad. Sci. U.S.A.} \textbf{2020}, \emph{117},
  3382--3387\relax
\mciteBstWouldAddEndPuncttrue
\mciteSetBstMidEndSepPunct{\mcitedefaultmidpunct}
{\mcitedefaultendpunct}{\mcitedefaultseppunct}\relax
\EndOfBibitem
\bibitem[Li \latin{et~al.}(2012)Li, Kr\"oger, and Liu]{LiYing2012}
Li,~Y.; Kr\"oger,~M.; Liu,~W.~K. Nanoparticle Effect on the Dynamics of Polymer
  Chains and Their Entanglement Network. \emph{Phys. Rev. Lett.} \textbf{2012},
  \emph{109}, 118001\relax
\mciteBstWouldAddEndPuncttrue
\mciteSetBstMidEndSepPunct{\mcitedefaultmidpunct}
{\mcitedefaultendpunct}{\mcitedefaultseppunct}\relax
\EndOfBibitem
\bibitem[Chen \latin{et~al.}(2019)Chen, Zhao, Shi, Lin, Jia, Qian, Lu, Zhang,
  Li, and Sun]{Chen2019}
Chen,~T.; Zhao,~H.-Y.; Shi,~R.; Lin,~W.-F.; Jia,~X.-M.; Qian,~H.-J.; Lu,~Z.-Y.;
  Zhang,~X.-X.; Li,~Y.-K.; Sun,~Z.-Y. An unexpected N-dependence in the
  viscosity reduction in all-polymer nanocomposite. \emph{Nat. Commun.}
  \textbf{2019}, \emph{10}, 5552\relax
\mciteBstWouldAddEndPuncttrue
\mciteSetBstMidEndSepPunct{\mcitedefaultmidpunct}
{\mcitedefaultendpunct}{\mcitedefaultseppunct}\relax
\EndOfBibitem
\bibitem[Cao \latin{et~al.}(2019)Cao, Merlitz, and Forest]{Cao2019}
Cao,~X.-Z.; Merlitz,~H.; Forest,~M.~G. Nanoparticle Loading of Unentangled
  Polymers Induces Entanglement-Like Relaxation Modes and a Broad Sol–Gel
  Transition. \emph{J. Phys. Chem. Lett.} \textbf{2019}, \emph{10},
  4968--4973\relax
\mciteBstWouldAddEndPuncttrue
\mciteSetBstMidEndSepPunct{\mcitedefaultmidpunct}
{\mcitedefaultendpunct}{\mcitedefaultseppunct}\relax
\EndOfBibitem
\bibitem[Zhang \latin{et~al.}(2021)Zhang, Li, Hu, and Liu]{Zhang2021}
Zhang,~B.; Li,~J.; Hu,~J.; Liu,~L. Theory of polymer diffusion in
  polymer–nanoparticle mixtures: effect of nanoparticle concentration and
  polymer length. \emph{Soft Matter} \textbf{2021}, \emph{17}, 4632--4642\relax
\mciteBstWouldAddEndPuncttrue
\mciteSetBstMidEndSepPunct{\mcitedefaultmidpunct}
{\mcitedefaultendpunct}{\mcitedefaultseppunct}\relax
\EndOfBibitem
\bibitem[Slimani \latin{et~al.}(2014)Slimani, Bacova, Bernabei, Narros, Likos,
  and Moreno]{Moreno2014}
Slimani,~M.~Z.; Bacova,~P.; Bernabei,~M.; Narros,~A.; Likos,~C.~N.;
  Moreno,~A.~J. Cluster Glasses of Semiflexible Ring Polymers. \emph{ACS Macro.
  Lett.} \textbf{2014}, \emph{3}, 611--616\relax
\mciteBstWouldAddEndPuncttrue
\mciteSetBstMidEndSepPunct{\mcitedefaultmidpunct}
{\mcitedefaultendpunct}{\mcitedefaultseppunct}\relax
\EndOfBibitem
\bibitem[Pipertzis \latin{et~al.}(2018)Pipertzis, Hossain, Monteiro, and
  Floudas]{Floudas2018}
Pipertzis,~A.; Hossain,~M.~D.; Monteiro,~M.~J.; Floudas,~G. Segmental Dynamics
  in Multicyclic Polystyrenes. \emph{Macromolecules} \textbf{2018}, \emph{51},
  1488--1497\relax
\mciteBstWouldAddEndPuncttrue
\mciteSetBstMidEndSepPunct{\mcitedefaultmidpunct}
{\mcitedefaultendpunct}{\mcitedefaultseppunct}\relax
\EndOfBibitem
\bibitem[Vargas-Lara \latin{et~al.}(2019)Vargas-Lara, Pazmiño~Betancourt, and
  Douglas]{JCP_Douglas_2019}
Vargas-Lara,~F.; Pazmiño~Betancourt,~B.~A.; Douglas,~J.~F. Influence of knot
  complexity on glass-formation in low molecular mass ring polymer melts.
  \emph{J. Chem. Phys.} \textbf{2019}, \emph{150}, 101103\relax
\mciteBstWouldAddEndPuncttrue
\mciteSetBstMidEndSepPunct{\mcitedefaultmidpunct}
{\mcitedefaultendpunct}{\mcitedefaultseppunct}\relax
\EndOfBibitem
\bibitem[Liu \latin{et~al.}(2019)Liu, Liu, Zhang, Du, Wesdemiotis, and
  Cheng]{Stephen2019}
Liu,~Y.; Liu,~G.; Zhang,~W.; Du,~C.; Wesdemiotis,~C.; Cheng,~S. Z.~D.
  Cooperative Soft-Cluster Glass in Giant Molecular Clusters.
  \emph{Macromolecules} \textbf{2019}, \emph{52}, 4341--4348\relax
\mciteBstWouldAddEndPuncttrue
\mciteSetBstMidEndSepPunct{\mcitedefaultmidpunct}
{\mcitedefaultendpunct}{\mcitedefaultseppunct}\relax
\EndOfBibitem
\bibitem[Zhu \latin{et~al.}(2022)Zhu, Luo, Zou, Ouyang, Ruan, Liu, and
  Liu]{Liu2022}
Zhu,~Y.; Luo,~J.; Zou,~Q.; Ouyang,~X.; Ruan,~Y.; Liu,~Y.; Liu,~G. Glassy
  feature in melts of 3-dimensional architectured polymer blends.
  \emph{Polymer} \textbf{2022}, \emph{238}, 124336\relax
\mciteBstWouldAddEndPuncttrue
\mciteSetBstMidEndSepPunct{\mcitedefaultmidpunct}
{\mcitedefaultendpunct}{\mcitedefaultseppunct}\relax
\EndOfBibitem
\bibitem[Lo and Turner(2013)Lo, and Turner]{Lo_2013}
Lo,~W.-C.; Turner,~M.~S. The topological glass in ring polymers. \emph{EPL}
  \textbf{2013}, \emph{102}, 58005\relax
\mciteBstWouldAddEndPuncttrue
\mciteSetBstMidEndSepPunct{\mcitedefaultmidpunct}
{\mcitedefaultendpunct}{\mcitedefaultseppunct}\relax
\EndOfBibitem
\bibitem[Michieletto and Turner(2016)Michieletto, and Turner]{Michieletto2016}
Michieletto,~D.; Turner,~M.~S. A topologically driven glass in ring polymers.
  \emph{Proc. Natl. Acad. Sci. U.S.A.} \textbf{2016}, \emph{113},
  5195--5200\relax
\mciteBstWouldAddEndPuncttrue
\mciteSetBstMidEndSepPunct{\mcitedefaultmidpunct}
{\mcitedefaultendpunct}{\mcitedefaultseppunct}\relax
\EndOfBibitem
\bibitem[Michieletto \latin{et~al.}(2017)Michieletto, Nahali, and
  Rosa]{Michieletto2017}
Michieletto,~D.; Nahali,~N.; Rosa,~A. Glassiness and Heterogeneous Dynamics in
  Dense Solutions of Ring Polymers. \emph{Phys. Rev. Lett.} \textbf{2017},
  \emph{119}, 197801\relax
\mciteBstWouldAddEndPuncttrue
\mciteSetBstMidEndSepPunct{\mcitedefaultmidpunct}
{\mcitedefaultendpunct}{\mcitedefaultseppunct}\relax
\EndOfBibitem
\bibitem[Orlandini and Micheletti(2021)Orlandini, and
  Micheletti]{Orlandini_2021}
Orlandini,~E.; Micheletti,~C. Topological and physical links in soft matter
  systems. \emph{J. Phys. Condens. Mat.} \textbf{2021}, \emph{34}, 013002\relax
\mciteBstWouldAddEndPuncttrue
\mciteSetBstMidEndSepPunct{\mcitedefaultmidpunct}
{\mcitedefaultendpunct}{\mcitedefaultseppunct}\relax
\EndOfBibitem
\bibitem[Smrek \latin{et~al.}(2020)Smrek, Chubak, Likos, and Kremer]{Smrek2020}
Smrek,~J.; Chubak,~I.; Likos,~C.~N.; Kremer,~K. Active topological glass.
  \emph{Nat. Commun.} \textbf{2020}, \emph{11}, 26\relax
\mciteBstWouldAddEndPuncttrue
\mciteSetBstMidEndSepPunct{\mcitedefaultmidpunct}
{\mcitedefaultendpunct}{\mcitedefaultseppunct}\relax
\EndOfBibitem
\bibitem[Chubak \latin{et~al.}(2020)Chubak, Likos, Kremer, and
  Smrek]{Smrek2020b}
Chubak,~I.; Likos,~C.~N.; Kremer,~K.; Smrek,~J. Emergence of active topological
  glass through directed chain dynamics and nonequilibrium phase segregation.
  \emph{Phys. Rev. Research} \textbf{2020}, \emph{2}, 043249\relax
\mciteBstWouldAddEndPuncttrue
\mciteSetBstMidEndSepPunct{\mcitedefaultmidpunct}
{\mcitedefaultendpunct}{\mcitedefaultseppunct}\relax
\EndOfBibitem
\bibitem[Chubak \latin{et~al.}(2022)Chubak, Pachong, Kremer, Likos, and
  Smrek]{Smrek2022}
Chubak,~I.; Pachong,~S.~M.; Kremer,~K.; Likos,~C.~N.; Smrek,~J. Active
  Topological Glass Confined within a Spherical Cavity. \emph{Macromolecules}
  \textbf{2022}, \emph{55}, 956--964\relax
\mciteBstWouldAddEndPuncttrue
\mciteSetBstMidEndSepPunct{\mcitedefaultmidpunct}
{\mcitedefaultendpunct}{\mcitedefaultseppunct}\relax
\EndOfBibitem
\bibitem[Roth(2016)]{Roth2016}
Roth,~C.~B. \emph{Polymer Glasses}; CRC Press, Boca Raton, 2016\relax
\mciteBstWouldAddEndPuncttrue
\mciteSetBstMidEndSepPunct{\mcitedefaultmidpunct}
{\mcitedefaultendpunct}{\mcitedefaultseppunct}\relax
\EndOfBibitem
\bibitem[Hart \latin{et~al.}(2021)Hart, Hertzog, Rauscher, Rawe, Tranquilli,
  and Rowan]{Hart2021Review}
Hart,~L.~F.; Hertzog,~J.~E.; Rauscher,~P.~M.; Rawe,~B.~W.; Tranquilli,~M.~M.;
  Rowan,~S.~J. Material properties and applications of mechanically interlocked
  polymers. \emph{Nat. Rev. Mat.} \textbf{2021}, \emph{6}, 508--530\relax
\mciteBstWouldAddEndPuncttrue
\mciteSetBstMidEndSepPunct{\mcitedefaultmidpunct}
{\mcitedefaultendpunct}{\mcitedefaultseppunct}\relax
\EndOfBibitem
\bibitem[Mena-Hernando and Pérez(2019)Mena-Hernando, and
  Pérez]{Emilio2019Review}
Mena-Hernando,~S.; Pérez,~E.~M. Mechanically interlocked materials. Rotaxanes
  and catenanes beyond the small molecule. \emph{Chem. Soc. Rev.}
  \textbf{2019}, \emph{48}, 5016--5032\relax
\mciteBstWouldAddEndPuncttrue
\mciteSetBstMidEndSepPunct{\mcitedefaultmidpunct}
{\mcitedefaultendpunct}{\mcitedefaultseppunct}\relax
\EndOfBibitem
\bibitem[Niu and Gibson(2009)Niu, and Gibson]{Niu2009}
Niu,~Z.; Gibson,~H.~W. Polycatenanes. \emph{Chem. Rev.} \textbf{2009},
  \emph{109}, 6024--6046\relax
\mciteBstWouldAddEndPuncttrue
\mciteSetBstMidEndSepPunct{\mcitedefaultmidpunct}
{\mcitedefaultendpunct}{\mcitedefaultseppunct}\relax
\EndOfBibitem
\bibitem[Arunachalam and Gibson(2014)Arunachalam, and Gibson]{Gibson2014}
Arunachalam,~M.; Gibson,~H.~W. Recent developments in polypseudorotaxanes and
  polyrotaxanes. \emph{Prog. Polym. Sci.} \textbf{2014}, \emph{39},
  1043--1073\relax
\mciteBstWouldAddEndPuncttrue
\mciteSetBstMidEndSepPunct{\mcitedefaultmidpunct}
{\mcitedefaultendpunct}{\mcitedefaultseppunct}\relax
\EndOfBibitem
\bibitem[Weidmann \latin{et~al.}(1999)Weidmann, Kern, Sauvage, Muscat, Mullins,
  Köhler, Rosenauer, Räder, Martin, and Geerts]{Weidmann1999}
Weidmann,~J.-L.; Kern,~J.-M.; Sauvage,~J.-P.; Muscat,~D.; Mullins,~S.;
  Köhler,~W.; Rosenauer,~C.; Räder,~H.~J.; Martin,~K.; Geerts,~Y.
  Poly[2]catenanes and Cyclic Oligo[2]catenanes Containing Alternating
  Topological and Covalent Bonds: Synthesis and Characterization.
  \emph{Chem-Eur. J.} \textbf{1999}, \emph{5}, 1841--1851\relax
\mciteBstWouldAddEndPuncttrue
\mciteSetBstMidEndSepPunct{\mcitedefaultmidpunct}
{\mcitedefaultendpunct}{\mcitedefaultseppunct}\relax
\EndOfBibitem
\bibitem[Chen \latin{et~al.}(2022)Chen, Sheng, Li, and Huang]{Chen2022}
Chen,~L.; Sheng,~X.; Li,~G.; Huang,~F. Mechanically interlocked polymers based
  on rotaxanes. \emph{Chem. Soc. Rev.} \textbf{2022}, \emph{51},
  7046--7065\relax
\mciteBstWouldAddEndPuncttrue
\mciteSetBstMidEndSepPunct{\mcitedefaultmidpunct}
{\mcitedefaultendpunct}{\mcitedefaultseppunct}\relax
\EndOfBibitem
\bibitem[Au-Yeung and Deng(2022)Au-Yeung, and Deng]{Yulin2022}
Au-Yeung,~H.~Y.; Deng,~Y. Distinctive features and challenges in catenane
  chemistry. \emph{Chem. Sci.} \textbf{2022}, \emph{13}, 3315--3334\relax
\mciteBstWouldAddEndPuncttrue
\mciteSetBstMidEndSepPunct{\mcitedefaultmidpunct}
{\mcitedefaultendpunct}{\mcitedefaultseppunct}\relax
\EndOfBibitem
\bibitem[Rauscher \latin{et~al.}(2018)Rauscher, Rowan, and
  de~Pablo]{Rauscher2018AML}
Rauscher,~P.~M.; Rowan,~S.~J.; de~Pablo,~J.~J. Topological Effects in Isolated
  Poly[n]catenanes: Molecular Dynamics Simulations and Rouse Mode Analysis.
  \emph{ACS Macro. Lett.} \textbf{2018}, \emph{7}, 938--943\relax
\mciteBstWouldAddEndPuncttrue
\mciteSetBstMidEndSepPunct{\mcitedefaultmidpunct}
{\mcitedefaultendpunct}{\mcitedefaultseppunct}\relax
\EndOfBibitem
\bibitem[Rauscher \latin{et~al.}(2020)Rauscher, Schweizer, Rowan, and
  de~Pablo]{Rauscher2020MA}
Rauscher,~P.~M.; Schweizer,~K.~S.; Rowan,~S.~J.; de~Pablo,~J.~J. Thermodynamics
  and Structure of Poly[n]catenane Melts. \emph{Macromolecules} \textbf{2020},
  \emph{53}, 3390--3408\relax
\mciteBstWouldAddEndPuncttrue
\mciteSetBstMidEndSepPunct{\mcitedefaultmidpunct}
{\mcitedefaultendpunct}{\mcitedefaultseppunct}\relax
\EndOfBibitem
\bibitem[Rauscher \latin{et~al.}(2020)Rauscher, Schweizer, Rowan, and
  de~Pablo]{Rauscher2020JCP}
Rauscher,~P.~M.; Schweizer,~K.~S.; Rowan,~S.~J.; de~Pablo,~J.~J. Dynamics of
  poly[n]catenane melts. \emph{J. Chem. Phys.} \textbf{2020}, \emph{152},
  214901\relax
\mciteBstWouldAddEndPuncttrue
\mciteSetBstMidEndSepPunct{\mcitedefaultmidpunct}
{\mcitedefaultendpunct}{\mcitedefaultseppunct}\relax
\EndOfBibitem
\bibitem[Hoyas~Pérez and Lewis(2020)Hoyas~Pérez, and Lewis]{Lewis2020}
Hoyas~Pérez,~N.; Lewis,~J. E.~M. Synthetic strategies towards mechanically
  interlocked oligomers and polymers. \emph{Org. Biomol. Chem.} \textbf{2020},
  \emph{18}, 6757--6780\relax
\mciteBstWouldAddEndPuncttrue
\mciteSetBstMidEndSepPunct{\mcitedefaultmidpunct}
{\mcitedefaultendpunct}{\mcitedefaultseppunct}\relax
\EndOfBibitem
\bibitem[Leigh \latin{et~al.}(2014)Leigh, Pritchard, and Stephens]{Leigh2014}
Leigh,~D.~A.; Pritchard,~R.~G.; Stephens,~A.~J. A Star of David catenane.
  \emph{Nat. Chem.} \textbf{2014}, \emph{6}, 978--982\relax
\mciteBstWouldAddEndPuncttrue
\mciteSetBstMidEndSepPunct{\mcitedefaultmidpunct}
{\mcitedefaultendpunct}{\mcitedefaultseppunct}\relax
\EndOfBibitem
\bibitem[Chichak \latin{et~al.}(2004)Chichak, Cantrill, Pease, Chiu, Cave,
  Atwood, and Stoddart]{Kelly2004}
Chichak,~K.~S.; Cantrill,~S.~J.; Pease,~A.~R.; Chiu,~S.-H.; Cave,~G. W.~V.;
  Atwood,~J.~L.; Stoddart,~J.~F. Molecular Borromean Rings. \emph{Science}
  \textbf{2004}, \emph{304}, 1308--1312\relax
\mciteBstWouldAddEndPuncttrue
\mciteSetBstMidEndSepPunct{\mcitedefaultmidpunct}
{\mcitedefaultendpunct}{\mcitedefaultseppunct}\relax
\EndOfBibitem
\bibitem[Danon \latin{et~al.}(2017)Danon, Krüger, Leigh, Lemonnier, Stephens,
  Vitorica-Yrezabal, and Woltering]{Danon2017}
Danon,~J.~J.; Krüger,~A.; Leigh,~D.~A.; Lemonnier,~J.-F.; Stephens,~A.~J.;
  Vitorica-Yrezabal,~I.~J.; Woltering,~S.~L. Braiding a molecular knot with
  eight crossings. \emph{Science} \textbf{2017}, \emph{355}, 159--162\relax
\mciteBstWouldAddEndPuncttrue
\mciteSetBstMidEndSepPunct{\mcitedefaultmidpunct}
{\mcitedefaultendpunct}{\mcitedefaultseppunct}\relax
\EndOfBibitem
\bibitem[Dietrich-Buchecker and Sauvage()Dietrich-Buchecker, and
  Sauvage]{Sauvage1989}
Dietrich-Buchecker,~C.~O.; Sauvage,~J.-P. A Synthetic Molecular Trefoil Knot.
  \emph{Angew. Chem. Int. Ed. Engl.} \emph{28}, 189--192\relax
\mciteBstWouldAddEndPuncttrue
\mciteSetBstMidEndSepPunct{\mcitedefaultmidpunct}
{\mcitedefaultendpunct}{\mcitedefaultseppunct}\relax
\EndOfBibitem
\bibitem[Marcos \latin{et~al.}(2016)Marcos, Stephens, Jaramillo-Garcia,
  Nussbaumer, Woltering, Valero, Lemonnier, Vitorica-Yrezabal, and
  Leigh]{Marcos2016}
Marcos,~V.; Stephens,~A.~J.; Jaramillo-Garcia,~J.; Nussbaumer,~A.~L.;
  Woltering,~S.~L.; Valero,~A.; Lemonnier,~J.-F.; Vitorica-Yrezabal,~I.~J.;
  Leigh,~D.~A. Allosteric initiation and regulation of catalysis with a
  molecular knot. \emph{Science} \textbf{2016}, \emph{352}, 1555--1559\relax
\mciteBstWouldAddEndPuncttrue
\mciteSetBstMidEndSepPunct{\mcitedefaultmidpunct}
{\mcitedefaultendpunct}{\mcitedefaultseppunct}\relax
\EndOfBibitem
\bibitem[Pazmi{\~n}o~Betancourt \latin{et~al.}(2015)Pazmi{\~n}o~Betancourt,
  Hanakata, Starr, and Douglas]{Douglas2015PNAS}
Pazmi{\~n}o~Betancourt,~B.~A.; Hanakata,~P.~Z.; Starr,~F.~W.; Douglas,~J.~F.
  Quantitative relations between cooperative motion, emergent elasticity, and
  free volume in model glass-forming polymer materials. \emph{Proc. Natl. Acad.
  Sci. U.S.A.} \textbf{2015}, \emph{112}, 2966--2971\relax
\mciteBstWouldAddEndPuncttrue
\mciteSetBstMidEndSepPunct{\mcitedefaultmidpunct}
{\mcitedefaultendpunct}{\mcitedefaultseppunct}\relax
\EndOfBibitem
\bibitem[Kremer and Grest(1990)Kremer, and Grest]{Kremer1990JCP}
Kremer,~K.; Grest,~G.~S. Dynamics of entangled linear polymer melts: A
  molecular‐dynamics simulation. \emph{J. Chem. Phys.} \textbf{1990},
  \emph{92}, 5057--5086\relax
\mciteBstWouldAddEndPuncttrue
\mciteSetBstMidEndSepPunct{\mcitedefaultmidpunct}
{\mcitedefaultendpunct}{\mcitedefaultseppunct}\relax
\EndOfBibitem
\bibitem[Bennemann \latin{et~al.}(1998)Bennemann, Paul, Binder, and
  D\"unweg]{Binder1998}
Bennemann,~C.; Paul,~W.; Binder,~K.; D\"unweg,~B. Molecular-dynamics
  simulations of the thermal glass transition in polymer melts:
  \ensuremath{\alpha}-relaxation behavior. \emph{Phys. Rev. E} \textbf{1998},
  \emph{57}, 843--851\relax
\mciteBstWouldAddEndPuncttrue
\mciteSetBstMidEndSepPunct{\mcitedefaultmidpunct}
{\mcitedefaultendpunct}{\mcitedefaultseppunct}\relax
\EndOfBibitem
\bibitem[Li and Zhang(2020)Li, and Zhang]{Li2020}
Li,~J.; Zhang,~B. Effect of chain length on structure and dynamics in a melt of
  semiflexible rings. \emph{EPL} \textbf{2020}, \emph{130}, 56001\relax
\mciteBstWouldAddEndPuncttrue
\mciteSetBstMidEndSepPunct{\mcitedefaultmidpunct}
{\mcitedefaultendpunct}{\mcitedefaultseppunct}\relax
\EndOfBibitem
\bibitem[Bernabei \latin{et~al.}(2013)Bernabei, Bacova, Moreno, Narros, and
  Likos]{Moreno2013}
Bernabei,~M.; Bacova,~P.; Moreno,~A.~J.; Narros,~A.; Likos,~C.~N. Fluids of
  semiflexible ring polymers: effective potentials and clustering. \emph{Soft
  Matter} \textbf{2013}, \emph{9}, 1287--1300\relax
\mciteBstWouldAddEndPuncttrue
\mciteSetBstMidEndSepPunct{\mcitedefaultmidpunct}
{\mcitedefaultendpunct}{\mcitedefaultseppunct}\relax
\EndOfBibitem
\bibitem[Ding \latin{et~al.}(2004)Ding, Kisliuk, and Sokolov]{Ding2004}
Ding,~Y.; Kisliuk,~A.; Sokolov,~A.~P. When Does a Molecule Become a Polymer?
  \emph{Macromolecules} \textbf{2004}, \emph{37}, 161--166\relax
\mciteBstWouldAddEndPuncttrue
\mciteSetBstMidEndSepPunct{\mcitedefaultmidpunct}
{\mcitedefaultendpunct}{\mcitedefaultseppunct}\relax
\EndOfBibitem
\bibitem[Chong \latin{et~al.}(2007)Chong, Aichele, Meyer, Fuchs, and
  Baschnagel]{Chong2007}
Chong,~S.-H.; Aichele,~M.; Meyer,~H.; Fuchs,~M.; Baschnagel,~J. Structural and
  conformational dynamics of supercooled polymer melts: Insights from
  first-principles theory and simulations. \emph{Phys. Rev. E} \textbf{2007},
  \emph{76}, 051806\relax
\mciteBstWouldAddEndPuncttrue
\mciteSetBstMidEndSepPunct{\mcitedefaultmidpunct}
{\mcitedefaultendpunct}{\mcitedefaultseppunct}\relax
\EndOfBibitem
\bibitem[Frey \latin{et~al.}(2015)Frey, Weysser, Meyer, Farago, Fuchs, and
  Baschnagel]{Frey2015}
Frey,~S.; Weysser,~F.; Meyer,~H.; Farago,~J.; Fuchs,~M.; Baschnagel,~J.
  Simulated glass-forming polymer melts: Dynamic scattering functions, chain
  length effects, and mode-coupling theory analysis. \emph{Eur. Phys. J. E}
  \textbf{2015}, \emph{38}, 11\relax
\mciteBstWouldAddEndPuncttrue
\mciteSetBstMidEndSepPunct{\mcitedefaultmidpunct}
{\mcitedefaultendpunct}{\mcitedefaultseppunct}\relax
\EndOfBibitem
\bibitem[Rubinstein and Colby(2003)Rubinstein, and Colby]{Rubinstein2003}
Rubinstein,~M.; Colby,~R.~H. \emph{Polymer physics}; Oxford University Press,
  2003\relax
\mciteBstWouldAddEndPuncttrue
\mciteSetBstMidEndSepPunct{\mcitedefaultmidpunct}
{\mcitedefaultendpunct}{\mcitedefaultseppunct}\relax
\EndOfBibitem
\bibitem[Bernabei \latin{et~al.}(2008)Bernabei, Moreno, and
  Colmenero]{Colmenero2008}
Bernabei,~M.; Moreno,~A.~J.; Colmenero,~J. Dynamic Arrest in Polymer Melts:
  Competition between Packing and Intramolecular Barriers. \emph{Phys. Rev.
  Lett.} \textbf{2008}, \emph{101}, 255701\relax
\mciteBstWouldAddEndPuncttrue
\mciteSetBstMidEndSepPunct{\mcitedefaultmidpunct}
{\mcitedefaultendpunct}{\mcitedefaultseppunct}\relax
\EndOfBibitem
\bibitem[Saltzman and Schweizer(2007)Saltzman, and Schweizer]{Saltzman2007}
Saltzman,~E.~J.; Schweizer,~K.~S. Short time properties, dynamic fragility and
  pressure effects in deeply supercooled polymer melts. \emph{J. Phys. Condens.
  Mat.} \textbf{2007}, \emph{19}, 205123\relax
\mciteBstWouldAddEndPuncttrue
\mciteSetBstMidEndSepPunct{\mcitedefaultmidpunct}
{\mcitedefaultendpunct}{\mcitedefaultseppunct}\relax
\EndOfBibitem
\bibitem[Chong \latin{et~al.}(2005)Chong, Moreno, Sciortino, and
  Kob]{Moreno2005}
Chong,~S.-H.; Moreno,~A.~J.; Sciortino,~F.; Kob,~W. Evidence for the Weak
  Steric Hindrance Scenario in the Supercooled-State Reorientational Dynamics.
  \emph{Phys. Rev. Lett.} \textbf{2005}, \emph{94}, 215701\relax
\mciteBstWouldAddEndPuncttrue
\mciteSetBstMidEndSepPunct{\mcitedefaultmidpunct}
{\mcitedefaultendpunct}{\mcitedefaultseppunct}\relax
\EndOfBibitem
\bibitem[Kob and Andersen(1994)Kob, and Andersen]{Kob1994}
Kob,~W.; Andersen,~H.~C. Scaling Behavior in the
  $\ensuremath{\beta}$-Relaxation Regime of a Supercooled Lennard-Jones
  Mixture. \emph{Phys. Rev. Lett.} \textbf{1994}, \emph{73}, 1376--1379\relax
\mciteBstWouldAddEndPuncttrue
\mciteSetBstMidEndSepPunct{\mcitedefaultmidpunct}
{\mcitedefaultendpunct}{\mcitedefaultseppunct}\relax
\EndOfBibitem
\bibitem[Sastry(2001)]{Sastry2001}
Sastry,~S. The relationship between fragility, configurational entropy and the
  potential energy landscape of glass-forming liquids. \emph{Nature}
  \textbf{2001}, \emph{409}, 164--167\relax
\mciteBstWouldAddEndPuncttrue
\mciteSetBstMidEndSepPunct{\mcitedefaultmidpunct}
{\mcitedefaultendpunct}{\mcitedefaultseppunct}\relax
\EndOfBibitem
\bibitem[McKenzie-Smith \latin{et~al.}(2021)McKenzie-Smith, Douglas, and
  Starr]{Douglas2021}
McKenzie-Smith,~T.~Q.; Douglas,~J.~F.; Starr,~F.~W. Explaining the Sensitivity
  of Polymer Segmental Relaxation to Additive Size Based on the Localization
  Model. \emph{Phys. Rev. Lett.} \textbf{2021}, \emph{127}, 277802\relax
\mciteBstWouldAddEndPuncttrue
\mciteSetBstMidEndSepPunct{\mcitedefaultmidpunct}
{\mcitedefaultendpunct}{\mcitedefaultseppunct}\relax
\EndOfBibitem
\bibitem[Kumar \latin{et~al.}(2013)Kumar, Goswami, Sumpter, Novikov, and
  Sokolov]{Kumar2013}
Kumar,~R.; Goswami,~M.; Sumpter,~B.~G.; Novikov,~V.~N.; Sokolov,~A.~P. Effects
  of backbone rigidity on the local structure and dynamics in polymer melts and
  glasses. \emph{Phys. Chem. Chem. Phys.} \textbf{2013}, \emph{15},
  4604--4609\relax
\mciteBstWouldAddEndPuncttrue
\mciteSetBstMidEndSepPunct{\mcitedefaultmidpunct}
{\mcitedefaultendpunct}{\mcitedefaultseppunct}\relax
\EndOfBibitem
\bibitem[Nguyen \latin{et~al.}(2015)Nguyen, Smith, Hoy, and
  Karayiannis]{Hong2015}
Nguyen,~H.~T.; Smith,~T.~B.; Hoy,~R.~S.; Karayiannis,~N.~C. Effect of chain
  stiffness on the competition between crystallization and glass-formation in
  model unentangled polymers. \emph{J. Chem. Phys.} \textbf{2015}, \emph{143},
  144901\relax
\mciteBstWouldAddEndPuncttrue
\mciteSetBstMidEndSepPunct{\mcitedefaultmidpunct}
{\mcitedefaultendpunct}{\mcitedefaultseppunct}\relax
\EndOfBibitem
\bibitem[Pan and Sun(2018)Pan, and Sun]{Sun2018}
Pan,~D.; Sun,~Z.-Y. Influence of chain stiffness on the dynamical heterogeneity
  and fragility of polymer melts. \emph{J. Chem. Phys.} \textbf{2018},
  \emph{149}, 234904\relax
\mciteBstWouldAddEndPuncttrue
\mciteSetBstMidEndSepPunct{\mcitedefaultmidpunct}
{\mcitedefaultendpunct}{\mcitedefaultseppunct}\relax
\EndOfBibitem
\bibitem[Bernabei \latin{et~al.}(2009)Bernabei, Moreno, and
  Colmenero]{Moreno2009}
Bernabei,~M.; Moreno,~A.~J.; Colmenero,~J. The role of intramolecular barriers
  on the glass transition of polymers: Computer simulations versus mode
  coupling theory. \emph{J. Chem. Phys.} \textbf{2009}, \emph{131},
  204502\relax
\mciteBstWouldAddEndPuncttrue
\mciteSetBstMidEndSepPunct{\mcitedefaultmidpunct}
{\mcitedefaultendpunct}{\mcitedefaultseppunct}\relax
\EndOfBibitem
\bibitem[Bernabei \latin{et~al.}(2011)Bernabei, Moreno, Zaccarelli, Sciortino,
  and Colmenero]{Moreno2011}
Bernabei,~M.; Moreno,~A.~J.; Zaccarelli,~E.; Sciortino,~F.; Colmenero,~J. From
  caging to Rouse dynamics in polymer melts with intramolecular barriers: A
  critical test of the mode coupling theory. \emph{J. Chem. Phys.}
  \textbf{2011}, \emph{134}, 024523\relax
\mciteBstWouldAddEndPuncttrue
\mciteSetBstMidEndSepPunct{\mcitedefaultmidpunct}
{\mcitedefaultendpunct}{\mcitedefaultseppunct}\relax
\EndOfBibitem
\bibitem[Colmenero(2015)]{Colmenero2015}
Colmenero,~J. Are polymers standard glass-forming systems? The role of
  intramolecular barriers on the glass-transition phenomena of glass-forming
  polymers. \emph{J. Phys. Condens. Mat.} \textbf{2015}, \emph{27},
  103101\relax
\mciteBstWouldAddEndPuncttrue
\mciteSetBstMidEndSepPunct{\mcitedefaultmidpunct}
{\mcitedefaultendpunct}{\mcitedefaultseppunct}\relax
\EndOfBibitem
\bibitem[Xu \latin{et~al.}(2020)Xu, Douglas, and Xu]{Xu2020}
Xu,~W.-S.; Douglas,~J.~F.; Xu,~X. Molecular Dynamics Study of Glass Formation
  in Polymer Melts with Varying Chain Stiffness. \emph{Macromolecules}
  \textbf{2020}, \emph{53}, 4796--4809\relax
\mciteBstWouldAddEndPuncttrue
\mciteSetBstMidEndSepPunct{\mcitedefaultmidpunct}
{\mcitedefaultendpunct}{\mcitedefaultseppunct}\relax
\EndOfBibitem
\bibitem[Schweizer and Saltzman(2004)Schweizer, and Saltzman]{Kenneth2004}
Schweizer,~K.~S.; Saltzman,~E.~J. Theory of dynamic barriers, activated
  hopping, and the glass transition in polymer melts. \emph{J. Chem. Phys.}
  \textbf{2004}, \emph{121}, 1984--2000\relax
\mciteBstWouldAddEndPuncttrue
\mciteSetBstMidEndSepPunct{\mcitedefaultmidpunct}
{\mcitedefaultendpunct}{\mcitedefaultseppunct}\relax
\EndOfBibitem
\bibitem[Pazmiño~Betancourt \latin{et~al.}(2014)Pazmiño~Betancourt, Douglas,
  and Starr]{Starr2014}
Pazmiño~Betancourt,~B.~A.; Douglas,~J.~F.; Starr,~F.~W. String model for the
  dynamics of glass-forming liquids. \emph{J. Chem. Phys.} \textbf{2014},
  \emph{140}, 204509\relax
\mciteBstWouldAddEndPuncttrue
\mciteSetBstMidEndSepPunct{\mcitedefaultmidpunct}
{\mcitedefaultendpunct}{\mcitedefaultseppunct}\relax
\EndOfBibitem
\bibitem[Sussman and Schweizer(2011)Sussman, and Schweizer]{Sussman2011}
Sussman,~D.~M.; Schweizer,~K.~S. Microscopic Theory of the Tube Confinement
  Potential for Liquids of Topologically Entangled Rigid Macromolecules.
  \emph{Phys. Rev. Lett.} \textbf{2011}, \emph{107}, 078102\relax
\mciteBstWouldAddEndPuncttrue
\mciteSetBstMidEndSepPunct{\mcitedefaultmidpunct}
{\mcitedefaultendpunct}{\mcitedefaultseppunct}\relax
\EndOfBibitem
\bibitem[Au-Yeung and Deng(2022)Au-Yeung, and Deng]{Deng2022}
Au-Yeung,~H.~Y.; Deng,~Y. Distinctive features and challenges in catenane
  chemistry. \emph{Chem. Sci.} \textbf{2022}, \emph{13}, 3315--3334\relax
\mciteBstWouldAddEndPuncttrue
\mciteSetBstMidEndSepPunct{\mcitedefaultmidpunct}
{\mcitedefaultendpunct}{\mcitedefaultseppunct}\relax
\EndOfBibitem
\bibitem[Zhang \latin{et~al.}(2022)Zhang, Zhao, Guo, Zhang, Pan, Wu, You, Yu,
  and Yan]{Zhang2022NC}
Zhang,~Z.; Zhao,~J.; Guo,~Z.; Zhang,~H.; Pan,~H.; Wu,~Q.; You,~W.; Yu,~W.;
  Yan,~X. Mechanically interlocked networks cross-linked by a molecular
  necklace. \emph{Nat. Commun.} \textbf{2022}, \emph{13}, 1393\relax
\mciteBstWouldAddEndPuncttrue
\mciteSetBstMidEndSepPunct{\mcitedefaultmidpunct}
{\mcitedefaultendpunct}{\mcitedefaultseppunct}\relax
\EndOfBibitem
\bibitem[Hagita \latin{et~al.}(2022)Hagita, Murashima, and Sakata]{Hagita2022}
Hagita,~K.; Murashima,~T.; Sakata,~N. Mathematical Classification and
  Rheological Properties of Ring Catenane Structures. \emph{Macromolecules}
  \textbf{2022}, \emph{55}, 166--177\relax
\mciteBstWouldAddEndPuncttrue
\mciteSetBstMidEndSepPunct{\mcitedefaultmidpunct}
{\mcitedefaultendpunct}{\mcitedefaultseppunct}\relax
\EndOfBibitem
\end{mcitethebibliography}

\clearpage 
\onecolumn

\Large For Table of Contents Use Only
\vspace{1cm}

	\includegraphics[width=8.25cm]{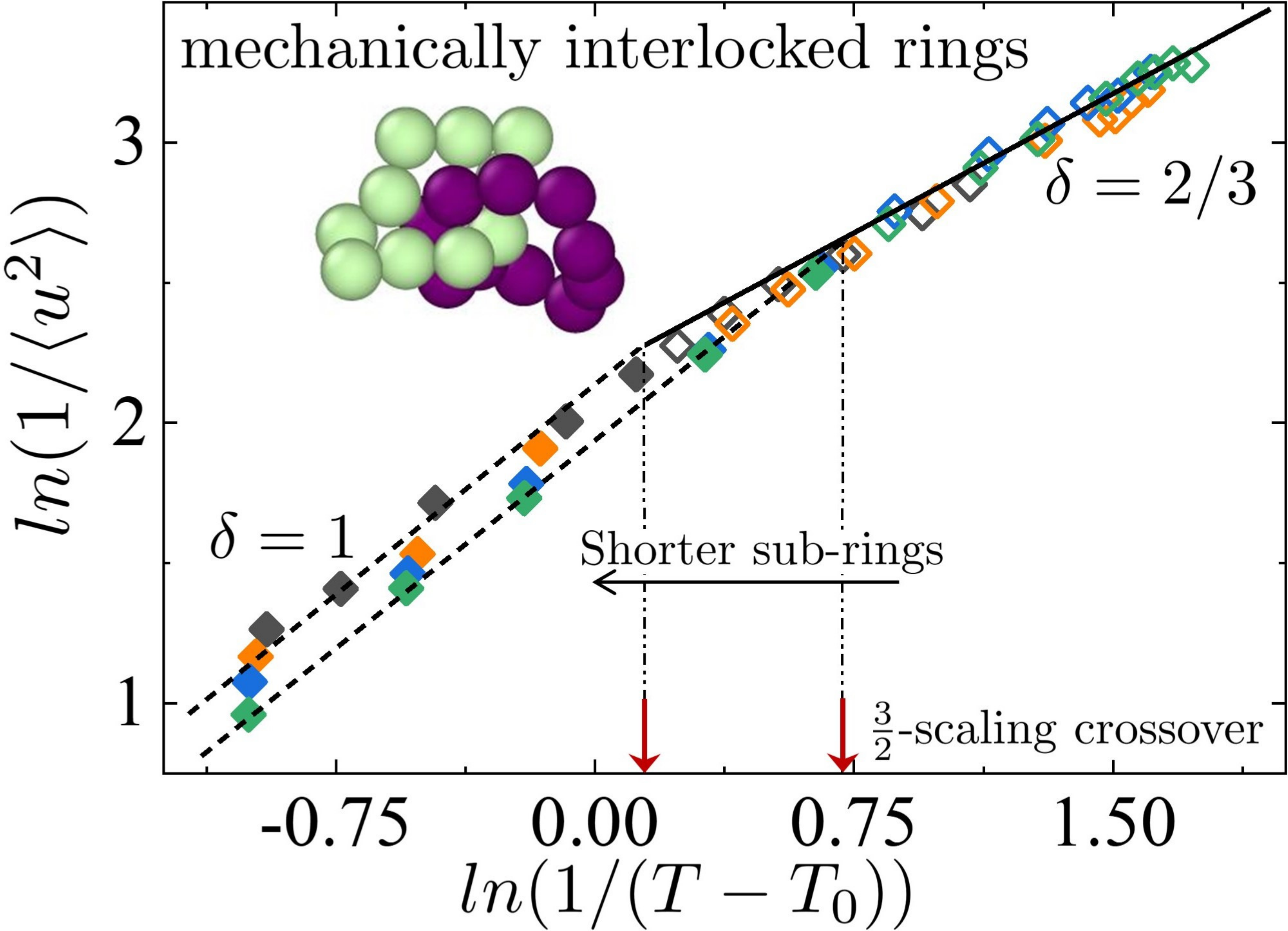}
	
\normalsize Glass Formation in Mechanically Interlocked Ring Polymers: the Effect of Chain Length and Topological Constraint
\vspace{1cm}

\normalsize Jian Li, Bokai Zhang, Yushan Li

\end{document}